\documentclass[aps,pra,reprint,onecolumn,groupedaddress,longbibliography]{revtex4-1}
\usepackage{comment}
\usepackage{lipsum}
\usepackage[version=3]{mhchem}
\usepackage{amsmath,bm}
\usepackage{amsfonts,amssymb}
\usepackage{dsfont}
\usepackage {setspace}
\usepackage{xcolor}
\usepackage{graphicx}
\graphicspath{figures/}
\usepackage[latin1]{inputenc}   
\usepackage[T1]{fontenc}      

\begin{document}
\title{Multispecies time-dependent restricted-active-space self-consistent-field theory for ultracold atomic and molecular gases}
\author{Camille L\'ev\^eque}
\affiliation{Department of Physics and Astronomy, Aarhus University, 8000 Aarhus C, Denmark}
\author{Lars Bojer Madsen}
\affiliation{Department of Physics and Astronomy, Aarhus University, 8000 Aarhus C, Denmark}

\date{\today}

\begin{abstract}
We discuss the multispecies time-dependent restricted-active-space self-consistent-field theory, an \textit{ab initio} wavefunction-based theory for mixtures of ultracold atomic and molecular gases. 
We present the general theory, based on the time-dependent variational principle, and derive the equations of motion. 
The theory captures in a time-dependent setting, via the specification of the restricted-active-space scheme, different levels of approximation from the mean-field to the full configuration interaction approach.
To assess its accuracy and to illustrate its ability to identify correlation effects at successive approximation levels, we apply the theory to compute the ground state energy of a Bose-Bose mixture interacting through a harmonic potential, for which the exact ground state energy is known analytically. We focus on the case of an ideal Bose gas interacting with a few impurities. The intra-species interaction between the impurities is relatively strong compared to the  inter-species interaction between the impurities and the ideal  noninteracting Bose gas.  
For this system, we find that an accurate description of the ground state necessitates 
the possibility of the theory to account for few-particle excitations out of the condensed phase;
a situation well accounted for by the present restricted-active-space theory for mixtures; 
and not within reach for approaches not incorporating orbital-restriction schemes.
\end{abstract}

\maketitle

\tableofcontents

\section{Introduction}

After the first realizations of Bose-Einstein condensates (BECs) \cite{Bradley95,Anderson95,Davis95} and condensation of fermionic atom pairs \cite{Regal04}, the experimental realization of Bose-Bose \cite{Myatt97,Hall98} and Bose-Fermi mixtures \cite{Ferrier14} has attracted much interest. Mixtures of cold atoms have lead to the opportunity to explore quantum effects such as the Kondo effect \cite{Gorshkov10}, quantum transport of impurity atoms \cite{Schirotzek09,Palzer09}, and to the possibility of realizing a single atom transistor \cite{Micheli04} and topological superfluids \cite{Midtgaard16,Midtgaard17}. A renewed interest has risen from the possibility to introduce impurities in a BEC, e.g.,  in collisionally induced transport in periodic potentials \cite{Ott04}. The possibility to trap a single \cite{Spethmann12} or multiple \cite{Santamore11,Scelle13} impurities in a BEC allows the investigation of the  physics of single and multiple polarons and their interactions \cite{Schirotzek09,Kohstall12,Koschorreck12,Zhang12,Christensen15,Levinsen15,Grusdt16,Hu16,Jorgensen16}.
The theoretical descriptions of mixtures of particles have often used a set of coupled Gross-Pitaevskii (GP) equations \cite{Uleysky14,Akram16_1,Akram16_2} or coupled hydrodynamic equations \cite{Wen17}, and thus described the system within the mean-field approximation. Recent works beyond the mean-field approximation include the use of the variational Lee-Low-Pines method \cite{Lee53} with a set of single-particle eigenstates for a single impurity \cite{Nakano17}, the consideration of quantized excitations of the ground state of a self-trapped impurity in a distorted Bose-Einstein condensate (BEC) \cite{Krzysztof06}, the Hubbard-Holstein model to describe impurities in an optical lattice immersed in a BEC \cite{Klein07} and Quantum Monte Carlo simulations \cite{Ceccarelli16}. 
For equilibrium and perturbative dynamics, theoretical approaches exist for beyond mean-field treatments \cite{Blume2001,Astrakharchik04} that could be extended to mixtures. In the time-dependent  case, and for non-perturbative dynamics induced, e.g., by a quench of the trap frequency \cite{Haller09,Fang14} or by particle interactions \cite{Lode12}, progress in theory  is challenged by the large Hilbert space associated with such dynamics. This holds in the case of  a single-component gas, i.e., in a gas containing a single species, and the problem is intensified  for multispecies mixtures as described below. In view of the increasing interest in time-dependent phenomena in mixtures of ultracold species, 
and the need for theory that accounts for correlations \cite{Trebbia06} and non-negligible depletion or fragmentation \cite{Streltsov04,Streltsov05,Alon05_1,Alon05_2,Mueller06,Brezinova12},
we explore in this work the possibility of formulating a time-dependent theory that can meet some of the challenges and still produce accurate results.

We focus on a wavefunction-based approach where  the many-body wavefunction is expanded in terms of a coherent superposition of configurations, with the latter build from \textit{time-dependent} orbitals that are optimally updated in each time-step. Without restrictions on the active orbital space, the multiconfigurational time-dependent Hartree (MCTDH) theory based on such time-dependent orbitals  was formulated for  nuclear dynamics in molecules \cite{Meyer90,Beck00}. Later the  MCTDH for fermions (MCTDHF) \cite{Zanghellini03} and the MCTDH for bosons (MCTDHB) \cite{Alon08} were formulated, as well as theory for mixtures \cite{Alon07,Cao13}. Eventhough the use of time-dependent orbitals reduces the number of configurations compared to approaches using time-independent orbitals \cite{Beck00}, these methods still suffer from computational challenges related to the exponential scaling of complexity with the number of particles, and application to situations with a large number of interacting particles is extremely challenging. The multi-layer extensions \cite{Wang03,Vendrell11,Cao13,Kronke13} allow consideration of more particles but do not fully resolve the problem with the increase in the number of configurations. 
Despite these limitations associated with the restriction on the number of particles, these methods have been extensively applied to address a whole range of processes in cold atom physics to describe few- to many-particle systems \cite{Schmitz13} in various dimensionalities \cite{Klaiman14,Bolsinger17}, trapping potentials \cite{Streltsov07,Lode17,Mistakidis17} and interaction regimes \cite{Zollner08,Streltsov11},
as summarized, e.g., in Ref.~\cite{Cao17}. 
Clearly, for applications, it would be desirable to be able to push these kinds of correlated time-dependent studies to particle numbers that are closer to those used in  experiments. The formulation of such a theory  is one of the aims of the present work.

The challenge with a large number of configurations is one of the reasons why recent related theory developments aiming at a time-dependent description of electrons in atoms and molecules \cite{Haru13,Sato13,Haru14_1,Haru14_2,Haxton15} have explored the possibility of introducing restrictions on the active orbital space by considering restricted-active-space (RAS) schemes. Very recently, the RAS idea was also explored in a time-dependent setting for single-species cold bosons  \cite{Leveque17}. 
The conclusion from these works is that the RAS, originally introduced in time-independent quantum chemistry \cite{Olsen88}, (i) can also in the time-dependent case significantly reduce the number of configurations and still give accurate results \cite{Haru13,Haru14_1,Leveque17} and (ii) can help identifying the configurations that are most important for capturing correlations involved in a  given dynamics \cite{Leveque17,Omiste17,Omiste18}. Without restrictions on the orbital excitations, the number of configurations is a particularly large challenge when mixtures of particles of different types are considered because of the multiconfigurational \textit{Ansatz} that includes the direct product space of the single-species configurations. In this work, we therefore formulate the multispecies time-dependent RAS self-consistent-field (TD-RASSCF) theory for mixtures of particles of different types. In this theory, we have control over the number of orbitals accessible for each particle type, as well as the excitation level. Thereby the number of configurations can be significantly reduced and the theory can be applied and address beyond mean-field effects for many more particles than is possible without incorporating RAS schemes.

Although the full potential of the present time-dependent theory will unfold in applications to nonequilibrium dynamics, we choose here, as a first application,  to focus on a critical ground-state test case, where the approach can be validated and the strength of the RAS schemes clearly illustrated.
We  focus on ground state energies for two-component Bose-Bose mixtures and validate the approach by comparison with a model where analytical results are available. 
We focus on the case, where a few relatively strongly interacting impurities interact with an ideal non-interacting Bose gas.
Using different RAS schemes, the role of correlation in this mixture can be comprehensively addressed. We show that while the number of orbitals plays a prominent role, the number of configurations only has a minor effect on the correlation energy. Specifically, for a given number of orbitals for the non-interacting bosons, the ground state energy converges with increasing excitation level of these bosons and with increasing number of orbitals for the impurities, described by fully correlated calculations, i.e., full configuration interaction (FCI) calculation. After certain cut-off values in excitation number and orbitals for the impurities, the accuracy does not improve, despite significant increase in the number of configurations. To converge to the exact analytical result, the number of orbitals for the non-interacting bosons has to be increased to a regime, where calculations based on the approach without the RAS, i.e., MCTDHB, would not be possible.
 This finding illustrates the value of incorporating the RAS concept, and physically it means that even for a small depletion of  a BEC, it is favorable for the particles out of the condensed orbital, i.e., out of the orbital with the highest occupation, to occupy higher energy orbitals separately rather than collectively. For the considered case of an ideal Bose gas interacting with a few impurities, we conclude that the mean-field description breaks down because of the interaction with the impurity atoms, and this irrespectively of the number of orbitals used to described these impurities. The small depletion mediated by the impurities has a large impact on the ground state energy and we show that both species must be described beyond the mean-field GP \textit{Ansatz} to sensitively reduce the error. 
Because the usual number of atoms in an experimental BEC is large, a wavefunction-based description beyond the mean-field approach is \textit{only} possible with a restriction on the number of configuration, i.e., with a RAS scheme.      

The paper is organized as follows. In Sec. II, we provide the theoretical background and a further motivation of the present approach. In Sec. III,  we derive the equations of motions from the time-dependent variational principle with emphasis on the specificities of mixtures. The generality of the theory is addressed, and it is discussed that, for specific choices of the RAS scheme,  it includes both  mean-field and full configuration interaction, multiconfigurational time-dependent Hartree  \textit{Ans\" atze} for fermions and bosons. In Sec. IV, we give an application illustrating the usefulness of the approach in obtaining accurate ground state energies with a reduced number of configurations, and we investigate the applicability range of the mean-field approach. Moreover, we show that the theory allows a comprehensive analysis of correlation in a given system and provides the relevant configurations that capture the correlation energy. In Sec. V, we summarize our main findings and conclude.

\section{Theoretical background}
\subsection{Many-body wavefunction} \label{mb_wf}
	The time evolution of a system composed of $N$ non-relativistic particles is governed by the time-dependent Schr\"odinger equation
\begin{equation}
i \frac{\partial}{\partial t}|\Psi(t)\rangle= H(t)|\Psi(t)\rangle,
\end{equation}
with $|\Psi(t)\rangle$ the time-dependent $N$-particle wavefunction. For a many-body system, the wavefunction has too many degrees of freedom to allow its simple expansion onto a grid. Thus, in wavefunction-based methods, approximations are used to obtain a numerically tractable expression for the wavefunction. For indistinguishable particles, a possible approximation is to use suitable linear combinations of time-dependent single-particle functions, or \textit{orbitals}. To take into account the statistics of the particles, the total wavefunction is expressed in terms of linear combinations of permanents for bosons and Slater determinants for fermions. Time-dependent orbitals provide a flexible description of the system as they adapt in space to accurately describe the evolution of the system. Thus fewer time-dependent orbitals are usually necessary in comparison to the use of time-independent orbitals. This smaller number of orbitals, in turn, considerably reduces the configurational space, i.e., reduces the number of all possible arrangements of the $N$ particles in all  the orbitals. In the wavefunction-based methodologies using time-dependent orbitals, such as, e.g., the MCTDH \cite{Meyer90}, the MCTDHB \cite{Alon08}, and the MCTDHF \cite{Zanghellini03} and their multi-layer extensions \cite{Wang03,Vendrell11,Cao13,Kronke13},  the orbitals span a time-evolving subspace, $\mathcal{P}$, of the single-particle Hilbert space, $\mathcal{H}$. The orthogonal complement, $\mathcal{Q}$, of the $\mathcal{P}$-space, collects all virtual orbitals, i.e., all orbitals that remain unoccupied in any of the configurations, such that $\mathcal{H}=\mathcal{P}+\mathcal{Q}$. During the time-evolution of the wavefunction, both $\mathcal{P}$ and $\mathcal{Q}$ spaces change in time, such that the projector $P(t)$ onto the $\mathcal{P}$-space is \textit{a priori} not time-invariant, i.e., $P(t_{1}) \ne P(t_{2})$ for $t_{1}\ne t_{2}$. The same relations holds for the projector $Q(t)$ onto the $\mathcal{Q}$-space.    

	In the following we describe a system of $K$ different kinds of particles. They form $K$ distinguishable groups each consisting of a specific type of  indistinguishable particles labeled by $\kappa$. In addition, the number of particles of each type, $N_{\kappa}$, is preserved during the time-evolution of the system, with the total number of particles, $N=\sum_{\kappa=1}^{K}N_{\kappa}$, being constant. Such a system has been described in the framework of the MCTDHB method \cite{Alon07} and its multi-layer extension \cite{Cao13,Kronke13}. We introduce a set of $M^{(\kappa)}$ time-dependent orbitals, $\{|\phi_{i_{\kappa}}^{(\kappa)}(t)\}$, for each particle type $\kappa = 1, 2, \dots,K$. Generalizing the 2-type Ansatz of Ref. \cite{Alon07}, to the case of $K$-types leads to the wavefunction,
\begin{eqnarray}\label{wf_FCI}
|\Psi(t)\rangle&=&
\sum_{\vec{n}_{1}\in \mathcal{V}_{\text{FCI}_{1}}}Ê\cdots \sum_{\vec{n}_{K}\in \mathcal{V}_{\text{FCI}_{K}}} 
{C_{\vec{n}_{1},\dots,\vec{n}_{K}}(t)|\vec{n}_{1},t}\rangle\otimes\cdots\otimes|\vec{n}_{K},t\rangle
\\ \nonumber
&\equiv& 
\sum_{\vec{n}_{1}\in \mathcal{V}_{\text{FCI}_{1}}}Ê\cdots \sum_{\vec{n}_{K}\in \mathcal{V}_{\text{FCI}_{K}}} 
{C_{\vec{n}_{1},\dots,\vec{n}_{K}}(t)|\vec{n}_{1},\dots,\vec{n}_{K},t}\rangle,
\end{eqnarray}
where $\vec{n}_{\kappa}$ denotes configurations of the particles of type $\kappa$. A given configuration $|\vec{n}_{\kappa},t\rangle$ in Eq. (\ref{wf_FCI}) is expanded in the orbitals $\{|\phi^{(\kappa)}_{i_{\kappa}}(t)\}$ taking the particle statistics into consideration. The indexes $\vec{n}_{\kappa}$ in the summations run over the full configurational subspace $\mathcal{V}_{\text{FCI}_\kappa}$ of each species, $\kappa$, as indicated by the subscript FCI. This formulation of the total wavefunction becomes exact when the number of orbitals becomes infinite. In practice only a finite number of orbitals can be used. The number of configurations, $\mathcal{N}_{c}$, increases exponentially with the number of orbitals and particles \cite{Leveque17}, and this scaling limits applications to a small number of particles and orbitals. For instance, a system consisting of $100$ identical bosons and being described with $5$ orbitals generates $\mathcal{N}_{c}\sim 4.6 \times 10^{6}$ configurations, a two-type mixture consisting of $13$ bosons and $5$ orbitals for each type generates $\mathcal{N}_{c}\sim 5.66 \times 10^{6}$ configurations, and a three-type mixture of $6$ bosons and $5$ orbitals of each type generates $\mathcal{N}_{c}\sim 9.26\times10^{6}$ configurations. Clearly, the number of configurations quickly becomes intractable using $\mathcal{V}_{\text{FCI}_{\kappa}}$, and in the next subsection we discuss how $\mathcal{N}_{c}$ can be reduced by imposing restrictions on the configurational space.      
   
\subsection{Restriction of the configurational space}   
   
	The TD-RASSCF method was recently introduced to reduce the exponential scaling of the configurational space. This method, first formulated for a system of electrons to study atomic and molecular systems \cite{Haru13,Haru14_1,Haru14_2}, was recently extended to a system of cold bosons \cite{Leveque17}. Here, we consider a further extension of the RAS with TD orbitals for a multispecies system composed of $K$ different types of particles, either bosons, fermions or mixtures.        

\begin{figure}
    \includegraphics[scale=0.5]{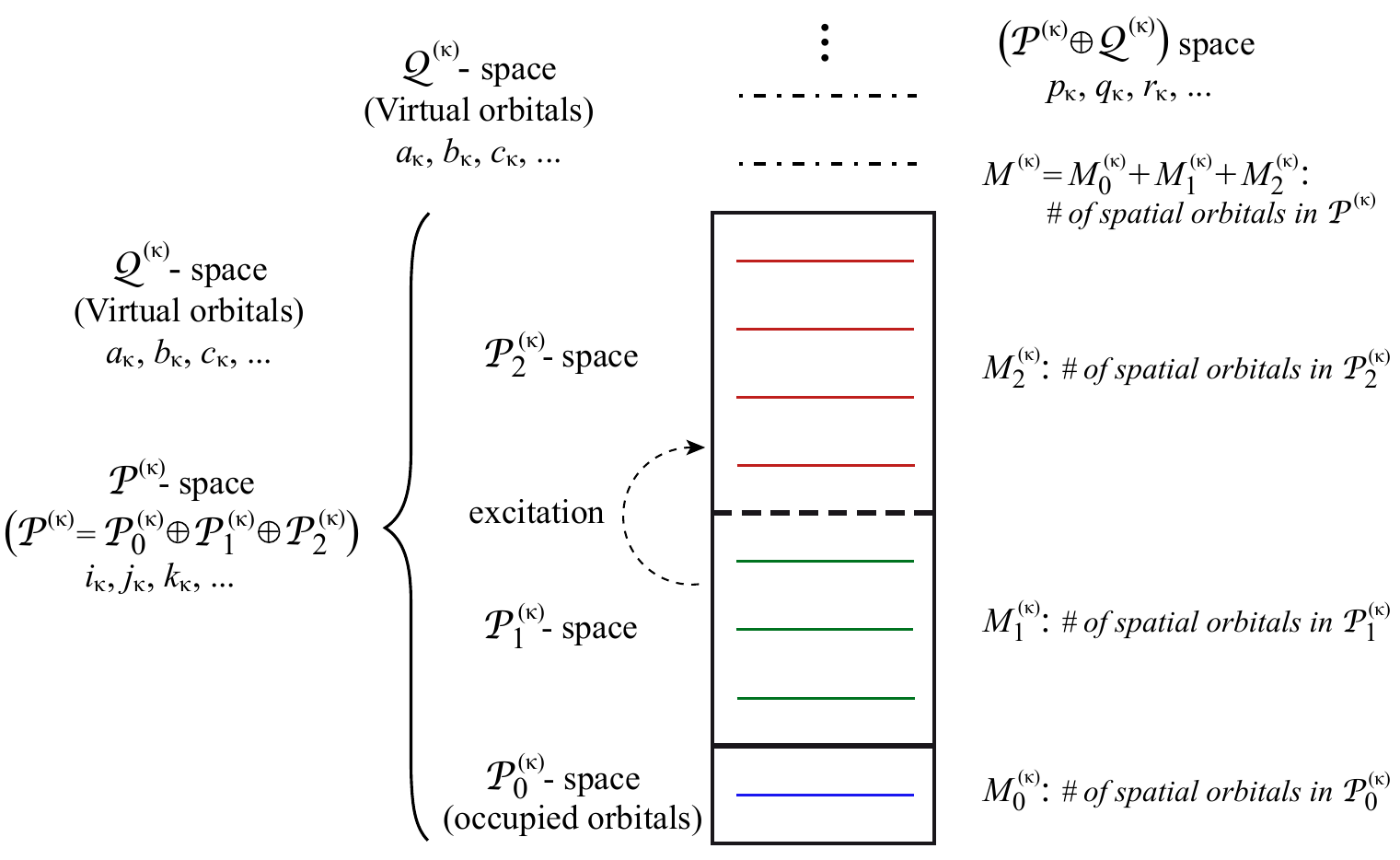}
\caption{Illustration of   the division of the single-particle Hilbert space, $\mathcal{H}^{(\kappa)}$ for the particles of type $\kappa$ in the multispecies TD-RASSCF theory and introduction of notations. The full space $\mathcal{H}^{(\kappa)}$ is spanned by the $\mathcal{P}^{(\kappa)}$ and $\mathcal{Q}^{(\kappa)}$ spaces, where the $\mathcal{Q}^{(\kappa)}$-space collects the unoccupied virtual orbitals (black, dashed-dotted lines in the top of the figure). The $M^{(\kappa)}$ occupied spatial orbitals span the $\mathcal{P}^{(\kappa)}$-space depicted by the rectangular (black) box.  The $\mathcal{P}^{(\kappa)}$-space is subdivided into three spaces namely, a core space, $\mathcal{P}^{(\kappa)}_{0}$, with $M^{(\kappa)}_{0}$ orbitals (blue line) that are always occupied, and two active spaces, with $M^{(\kappa)}_{1}$ orbitals (green lines) in the $\mathcal{P}^{(\kappa)}_{1}$-space and $M^{(\kappa)}_{2}$ orbitals (red lines) in the $\mathcal{P}^{(\kappa)}_{2}$-space. Excitations from the $\mathcal{P}^{(\kappa)}_{1}$-space to the $\mathcal{P}^{(\kappa)}_{2}$-space are subject to restrictions specified by the RAS scheme. The indexes used in this article are introduced in the figure, the orbitals in either $\mathcal{P}^{(\kappa)}$- or $\mathcal{Q}^{(\kappa)}$-space are labeled by $p_{\kappa}, q_{\kappa}, r_{\kappa}, \dots$. The $\mathcal{P}^{(\kappa)}$-space orbitals are labeled by $i_{\kappa}, j_{\kappa}, k_{\kappa}, \dots$, and the $\mathcal{Q}^{(\kappa)}$-space orbitals by $a_{\kappa}, b_{\kappa}, c_{\kappa},\dots$. In this illustration, there are 1, 3, and 4 spatial orbitals in $\mathcal{P}^{(\kappa)}_{0}$, $\mathcal{P}^{(\kappa)}_{1}$ and $\mathcal{P}^{(\kappa)}_{2}$ spaces, respectively.}   \label{Fig_1} 
 \end{figure}

The single-particle Hilbert space, $\mathcal{H}^{(\kappa)}$, for the particle of type $\kappa$ consists of a $\mathcal{P}^{(\kappa)}$-space spanned by the $M^{(\kappa)}$ time-dependent occupied orbitals and a  $\mathcal{Q}^{(\kappa)}$-space spanned by all virtual orbitals, see Fig. \ref{Fig_1}. In the TD-RASSCF theory, the $\mathcal{P}^{(\kappa)}$-space is divided into three different spaces, $\mathcal{P}_{0}^{(\kappa)}$, $\mathcal{P}_{1}^{(\kappa)}$ and $\mathcal{P}_{2}^{(\kappa)}$, such that $M_{0}^{(\kappa)}+M_{1}^{(\kappa)}+M_{2}^{(\kappa)} = M^{(\kappa)}$, with $M_{0}^{(\kappa)}$, $M_{1}^{(\kappa)}$ and $M_{2}^{(\kappa)}$ the number of orbitals in the $\mathcal{P}_{0}^{(\kappa)}$, $\mathcal{P}_{1}^{(\kappa)}$ and $\mathcal{P}_{2}^{(\kappa)}$ spaces, respectively. The $\mathcal{P}_{0}^{(\kappa)}$ space collects orbitals which are set to be fully occupied in all configurations. Eventhough the orbitals are time-dependent and optimally updated in each time step, the number of particles in these $\mathcal{P}_{0}^{(\kappa)}$ orbitals is hence constant in all the configurations. These ${\cal P}_0^{(\kappa)}$ orbitals are called inactive or core orbitals. The remaining spaces collect the active orbitals from which the configurations are generated according to a specified RAS scheme. The RAS scheme, in addition of specifying $M_{0}^{(\kappa)}$, $M_{1}^{(\kappa)}$ and $M_{2}^{(\kappa)}$, also specifies the number of excitations from $\mathcal{P}_{1}^{(\kappa)}$ to $\mathcal{P}_{2}^{(\kappa)}$  \cite{Olsen88}. For example, we can consider only single and double excitations (SD) from $\mathcal{P}_{1}^{(\kappa)}$ to $\mathcal{P}_{2}^{(\kappa)}$ to restrict the number of configurations. The $\mathcal{P}_{1}^{(\kappa)}$ space must include enough orbitals to accommodate all the particles of type $\kappa$, except the ones already described by the $\mathcal{P}_{0}^{(\kappa)}$ orbitals. All possible configurations of the particles in the $M_{1}^{(\kappa)}$ orbitals are included in the expansion of the total wavefunction. Then the restriction of the number of configurations, in addition to the $\mathcal{P}_{0}^{(\kappa)}$ orbitals, results from the specification of the number of particles that can be "excited", i.e., promoted, from the $\mathcal{P}_{1}^{(\kappa)}$ to the $\mathcal{P}_{2}^{(\kappa)}$ orbitals.

\begin{figure}[h!]
    \includegraphics[scale=0.5]{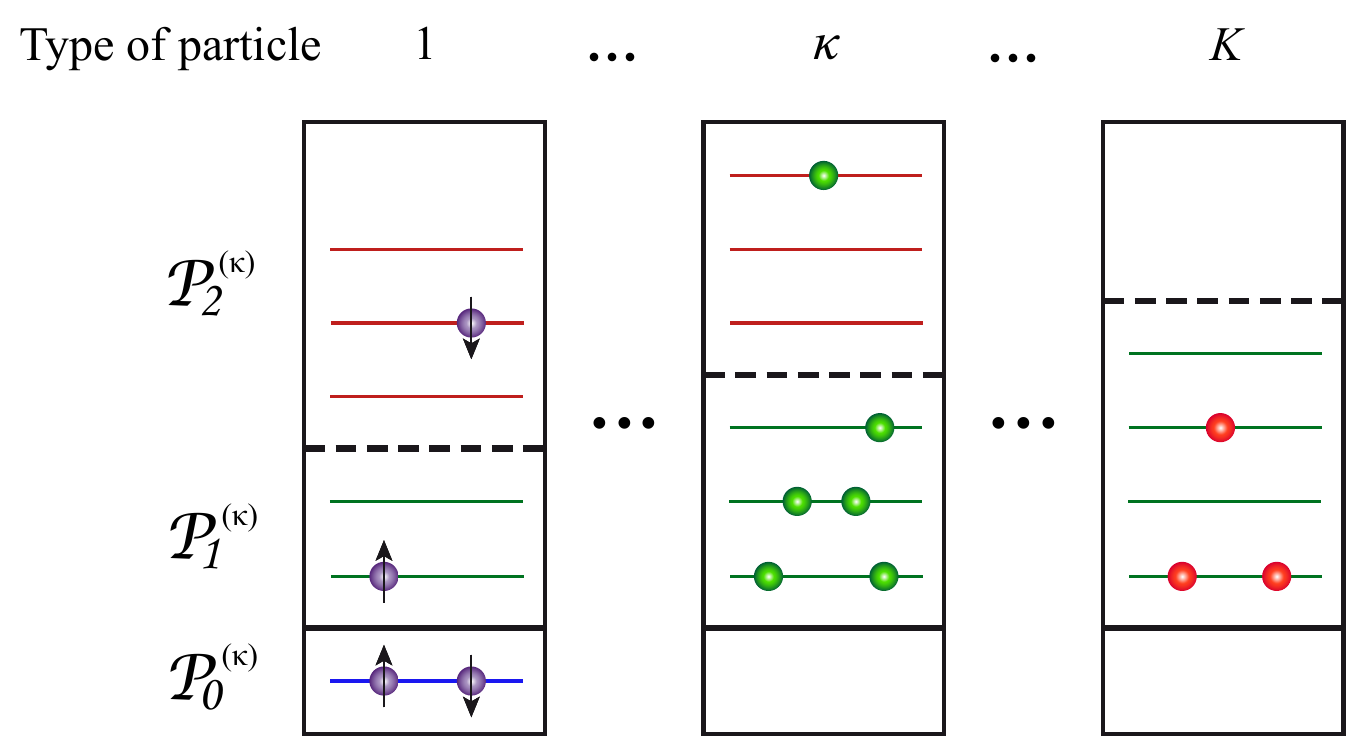}
    \caption{Illustration of a configuration contributing to the wavefunction of the multispecies TD-RASSCF theory for a system of $K$ different types of particles. In this example, there are 4 fermions of type 1 [(purple) full circles], 6 bosons of type $\kappa$ [(green) full circles] and 3 bosons of type $K$ [(red) full circles]. For  each species a specific level of theory can be used. For instance, in the realization in the figure, the fermions of type 1 are described using $M^{(1)}_{0}=1$ spatial orbital in $\mathcal{P}^{(1)}_{0}$ (blue line), $M^{(1)}_{1}=2$ spatial orbitals in $\mathcal{P}^{(1)}_{1}$ (green lines) and $M^{(1)}_{2}=3$ spatial orbitals in $\mathcal{P}^{(1)}_{2}$ (red lines). The bosons of type $\kappa$ are describes using $M^{(\kappa)}_{1}=3$ spatial orbitals in $\mathcal{P}^{(\kappa)}_{1}$ and $M^{(\kappa)}_{2}=3$ spatial orbitals in $\mathcal{P}^{(\kappa)}_{2}$ and the bosons of type $K$ by using $M^{(K)}_{1}=4$ spatial orbitals in $\mathcal{P}^{(K)}_{1}$, i.e., using an FCI MCTDHB wavefunction. The choice of the restrictions of the configurational spaces of each type of particles should be motivated by physical properties of the system at hand. }   \label{Fig_2} 
 \end{figure}

	Using the mentioned restrictions to limit the size of the configurational space for each type of particle, the multispecies TD-RASSCF \textit{Ansatz} reads, 
\begin{eqnarray}\label{RAS_wf}
|\Psi(t)\rangle&=&\sum_{\vec{n}_{1}\in \mathcal{V}_{\text{RAS}_{1}}}Ê\cdots \sum_{\vec{n}_{K}\in \mathcal{V}_{\text{RAS}_{K}}} C_{\vec{n}_{1},\dots,\vec{n}_{K}}(t)|\vec{n}_{1},\dots,\vec{n}_{K},t\rangle \\ \nonumber
&\equiv& 
\sum_{\vec{n}_{1},\dots,\vec{n}_{K}\in \mathcal{V}_\text{RAS}} 
C_{\vec{n}_{1},\dots,\vec{n}_{K}}(t)|\vec{n}_{1},\dots,\vec{n}_{K},t\rangle,
\end{eqnarray}
where we introduced a short-hand notation for the summations over configurations in the second line.
Figure \ref{Fig_2} illustrates in an artistic manner one of the possible configurations, i.e., one of the realization of $|\vec{n}_{1},\dots,\vec{n}_{K},t\rangle$. The strength of the \textit{Ansatz} of Eq. (\ref{RAS_wf}) is its generality, see Fig. \ref{Fig_2}, as it includes as limiting cases the FCI wavefunction, Eq. (\ref{wf_FCI}), as well as the mean-field wavefunction, when each type of particle is described by the GP \cite{Gross61,Pitaevskii61} or the Hartree-Fock (HF)  \cite{Kulander87} \textit{Ansatz}. Moreover, when $K=1$, Eq. (\ref{RAS_wf}), boils down to the TD-RASSCF theory for bosons \cite{Leveque17} or fermions \cite{Haru13,Haru14_1} including, as limiting cases, the MCTDH-F \cite{Zanghellini03} and -B \cite{Alon08} wavefunctions, the TD-CASSCF theory for fermions \cite{Sato13} where only $\mathcal{P}_{0}$ and $\mathcal{P}_{1}$ spaces are included, and the TDHF and TDGP \textit{Ansatz} for only fully occupied orbitals and a single orbital, respectively \cite{Leveque17}. Thus, Eq. (\ref{RAS_wf}) shows that by varying the number of orbitals in the different $\mathcal{P}_{i}^{(\kappa)}$ $(i=0,1,2)$ spaces and the allowed excitations from the $\mathcal{P}_{1}^{(\kappa)}$ to the $\mathcal{P}_{2}^{(\kappa)}$ space, we obtain a hierarchy in the accuracy of the description of the correlation between the particles, covering descriptions  from the less accurate mean-field \textit{Ansatz} to the most accurate FCI wavefunction. For mixtures, the number of configurations is a challenge, as discussed in Sec. \ref{mb_wf}. The multispecies TD-RASSCF wavefunction uses a restricted configurational space, $\mathcal{V}_{\text{RAS}_{\kappa}}$, for each particle type $\kappa$, and provides in this way a reduction of the number of configurations for the total wavefunction. The choice of the restrictions can be motivated by physical insight of the system at hand. For instance by consideration of the interaction strength between the particles or the trapping potentials. Moreover, for a single type of particle, the TD-RASSCF methods have shown that accurate results can be obtained with a substantially reduced number of configurations for both bosons and fermions, and with a clear hierarchy of accuracy~\cite{Haru13,Haru14_1,Leveque17}. For these reasons the multispecies TD-RASSCF theory seems to be suitable to tackle the exponential growing of the configurational space in mixtures. 

	With the wavefunction \textit{Ansatz} of Eq. (\ref{RAS_wf}) at hand, the equations of motion (EOM) can be derived. The EOM provide the time-evolution of the parameters entering the expression of wavefunction, i.e., the amplitudes $\{C_{\vec{n}_{1},\dots,\vec{n}_{K}}(t)\}$ and the orbitals $\{ |\phi^{(\kappa)}_{i_{\kappa}}(t)\rangle\}$. The time-evolution is obtained through the time derivative of the parameters, $\{\dot{C}_{\vec{n}_{1},\dots,\vec{n}_{K}}(t)\}$ and $\{ |\dot{\phi}^{(\kappa)}_{i_{\kappa}}(t)\rangle\}$, where the dot on top of a symbol denotes its time derivative. The time derivative of the orbitals can be expressed in terms of $\mathcal{P}^{(\kappa)}$- and $\mathcal{Q}^{(\kappa)}$-space contributions, using their respective projectors, $P^{(\kappa)}(t)$ and $Q^{(\kappa)}(t)$, 
\begin{equation}\label{deriv_orb}
|\dot{\phi}^{(\kappa)}_{i_{\kappa}}(t)\rangle = P^{(\kappa)}(t)|\dot{\phi}^{(\kappa)}_{i_{\kappa}}(t)\rangle + Q^{(\kappa)}(t)|\dot{\phi}^{(\kappa)}_{i_{\kappa}}(t)\rangle,
\end{equation}	
where the relation $P^{(\kappa)}(t)+Q^{(\kappa)}(t)=\mathds{1}$, with $\mathds{1}$ the identity operator, is used. The term, $P^{(\kappa)}(t)|\dot{\phi}^{(\kappa)}_{i_{\kappa}}(t)\rangle$, describes rotations of the orbitals of the $\mathcal{P}^{(\kappa)}$-space into each other. 
In the TD-RASSCF theories these rotations play a major role as the orbitals of the different $\mathcal{P}_{i}^{(\kappa)}$ (i=0,1,2) spaces are not equivalent because of the restriction on the configurations. From the second term, $Q^{(\kappa)}(t)|\dot{\phi}^{(\kappa)}_{i_{\kappa}}(t)\rangle$, we can understand why time-dependent orbitals are so efficient. Indeed, this term describes the evolution of the $\mathcal{P}^{(\kappa)}$-space orbitals toward the orthogonal $\mathcal{Q}^{(\kappa)}$-space. Thus, for $Q^{(\kappa)}(t)|\dot{\phi}^{(\kappa)}_{i_{\kappa}}(t)\rangle \ne 0$ the $\mathcal{P}^{(\kappa)}$-space orbitals dynamically span the relevant part of the single-particle Hilbert space. Note that if $Q^{(\kappa)}(t)|\dot{\phi}^{(\kappa)}_{i_{\kappa}}(t)\rangle=0$, then the time-evolution of the $\mathcal{P}^{(\kappa)}$-space orbitals reduces to rotations of the orbitals of that space into each other, and the single-particle Hilbert space spanned by the $\mathcal{P}^{(\kappa)}$-space orbitals is time-independent. The knowledge of the time derivative of the parameters allows the evaluation, using numerical integration schemes, of the time evolution of the total wavefuntion. In  the next section (Sec. \ref{Theo_frame}), we provide in detail the derivation of the EOM for the multispecies TD-RASSCF theory.

\section{Derivation of the equations of motion}\label{Theo_frame}
To derive the EOM for the set of amplitudes $\{C_{\vec{n}_{1},\dots,\vec{n}_{K}}(t)\}$ and orbitals $\{ |\phi^{(\kappa)}_{i_{\kappa}}(t)\rangle\}$ ($\kappa=1,\dots,K$) we use the time-dependent variational principle \cite{Dirac30,Frenkel34,McLachlan64,Lubich08}, with an action functional,
\begin{multline} 
S \left[ \{C_{\vec{n}_{1},\dots,\vec{n}_{K}}(t)\}, \{|\phi^{(\kappa)}_{i_{\kappa}}(t)\rangle\}_{\kappa=1}^K,\{\epsilon_{j_{\kappa}}^{i_{\kappa}(\kappa)}(t)\}_{\kappa=1}^K \right] =  \\
 \int_{0}^{T}\left[ \langle\Psi(t)| \left( i\partial_{t}-H(t)\right) |\Psi(t)\rangle  + \sum_{\kappa=1}^{K}\sum_{i_{\kappa}j_{\kappa}=1}^{M^{(\kappa)}}\epsilon_{j_{\kappa}}^{i_{\kappa}(\kappa)}(t)\left( \langle\phi_{i_{\kappa}}^{(\kappa)}(t)|\phi_{j_{\kappa}}^{(\kappa)}(t)\rangle - \delta_{j_{\kappa}i_{\kappa}} \right) \right] dt,
\end{multline}
where $\partial_{t}$ denotes the derivative with respect to time and the Lagrange multipliers, $\epsilon_{j_{\kappa}}^{i_{\kappa}({\kappa})}(t)$, ensure that the orbitals for the particles of type $\kappa$ remain orthonormal to each other at all $t$. Note that we do not need to ensure the orthonormality between the orbitals of different species. In the following, the orbitals that belong to the $\mathcal{P}^{(\kappa)}$-space are labelled by $i_{\kappa}, j_{\kappa}, k_{\kappa}, \dots$ and the orbitals that belong to the $\mathcal{Q}^{(\kappa)}$-space by $a_{\kappa}, b_{\kappa}, c_{\kappa}, \dots$ The labels $p_{\kappa}, q_{\kappa}, r_{\kappa}, s_{\kappa}, \dots$ are used for orbitals that belong to either the $\mathcal{P}^{(\kappa)}$-space or the $\mathcal{Q}^{(\kappa)}$-space, see Fig. \ref{Fig_1}. The total Hamiltonian of the system, $H(t)$, consists of the Hamiltonians describing each type of particle, $H_{\kappa}(t)$, and the pairwise interactions between any two different types of particles, $H_{\kappa\gamma}(t)$, such that,
\begin{equation}\label{H_tot}
H(t) = \sum_{\kappa=1}^{K} H_{\kappa}(t) + \frac{1}{2}\sum_{\kappa=1}^{K}\sum_{\gamma\ne\kappa}^{K} H_{\kappa\gamma}(t). 
\end{equation} 
We consider one- and two-body operators in the derivation of the EOM. We can write explicitly each term of Eq. (\ref{H_tot}) in the framework of second quantization, 
\begin{eqnarray}
H_{\kappa}(t)&=&\sum_{p_{\kappa}q_{\kappa}}h_{q_{\kappa}}^{p_{\kappa}(\kappa)}(t)b_{p_{\kappa}}^{\dag(\kappa)}b^{(\kappa)}_{q_{\kappa}}+\frac{1}{2}\sum_{p_{\kappa}q_{\kappa}r_{\kappa}s_{\kappa}}{v_{q_{\kappa}s_{\kappa}}^{p_{\kappa}r_{\kappa}(\kappa)}(t) b_{p_{\kappa}}^{\dag(\kappa)}b_{r_{\kappa}}^{\dag(\kappa)}b^{(\kappa)}_{s_{\kappa}}b^{(\kappa)}_{q_{\kappa}}}, \label{H_intra}  \\ 
H_{\kappa\gamma}(t)&=&\sum_{p_{\kappa}p_{\gamma}q_{\kappa}q_{\gamma}}{w_{q_{\kappa}q_{\gamma}}^{p_{\kappa}p_{\gamma}(\kappa\gamma)} (t)b_{p_{\kappa}}^{\dag(\kappa)}b^{(\kappa)}_{q_{\kappa}}b_{p_{\gamma}}^{\dag(\gamma)}b^{(\gamma)}_{q_{\gamma}}},  \label{H_inter}
\end{eqnarray}
where $b^{(\kappa)}_{p_{\kappa}}$ is the annihilation operator of a particle of type $\kappa$ in the spin-orbital $|\phi_{p_{\kappa}}^{(\kappa)}(t)\rangle$ and $b_{p_{\kappa}}^{\dag(\kappa)}$ the corresponding creation operator. These operators satisfy the usual commutation (anti-commutation) relations, $[ b^{(\kappa)}_{p_{\kappa}}, b_{q_{\kappa}}^{\dag(\kappa)} ] = b^{\kappa}_{p_{\kappa}}b_{q_{\kappa}}^{\dag(\kappa)}- b_{q_{\kappa}}^{\dag(\kappa)}b^{(\kappa)}_{p_{\kappa}} = \delta_{p_{\kappa}q_{\kappa}}$ $(\{ b^{(\kappa)}_{p_{\kappa}},b_{q_{\kappa}}^{\dag(\kappa)} \}= b^{\kappa}_{p_{\kappa}}b_{q_{\kappa}}^{\dag(\kappa)}+ b_{q_{\kappa}}^{\dag(\kappa)}b^{(\kappa)}_{p_{\kappa}} = \delta_{p_{\kappa}q_{\kappa}})$ for bosons (fermions). The matrix elements of the one- and two-body operators, in the basis of the time-dependent orbitals of each species, are expressed as,
\begin{eqnarray} 
h_{q_{\kappa}}^{p_{\kappa}(\kappa)}(t)&=&\int \phi_{p_{\kappa}}^{*(\kappa)}(\textbf{x}_{\kappa},t) h^{(\kappa)}(\textbf{x}_{\kappa},t) \phi_{q_{\kappa}}(\textbf{x}_{\kappa},t) d\textbf{x}_{\kappa}, \\
v_{q_{\kappa}s_{\kappa}}^{p_{\kappa}r_{\kappa}(\kappa)}(t) &=& \int \int \phi_{p_{\kappa}}^{*(\kappa)} (\textbf{x}_{\kappa},t) \phi_{r_{\kappa}}^{*(\kappa)}(\textbf{x}'_{\kappa},t) v^{(\kappa)}(\textbf{x}_{\kappa},\textbf{x}'_{\kappa},t) \phi^{(\kappa)}_{q_{\kappa}}(\textbf{x}_{\kappa},t)\phi^{(\kappa)}_{s_{\kappa}}(\textbf{x}'_{\kappa},t) d\textbf{x}_{\kappa}d\textbf{x}'_{\kappa},\\
w_{q_{\kappa} q_{\gamma}}^{p_{\kappa} p_{\gamma}(\kappa\gamma)} (t)&=& \int \int \phi_{p_{\kappa}}^{*(\kappa)} (\textbf{x}_{\kappa},t) \phi_{p_{\gamma}}^{*(\gamma)}(\textbf{x}_{\gamma},t) w^{(\kappa\gamma)}(\textbf{x}_{\kappa},\textbf{x}_{\gamma},t) \phi^{(\kappa)}_{q_{\kappa}}(\textbf{x}_{\kappa},t)\phi^{(\gamma)}_{q_{\gamma}}(\textbf{x}_{\gamma},t) d\textbf{x}_{\kappa}d\textbf{x}_{\gamma},
\end{eqnarray}
with $\textbf{x}_{\kappa}$ denoting generalized coordinates, including space and spin variables, for the particles of type $\kappa$. Note that even if the operators are time-independent, the matrix elements are time-dependent due to the use of time-dependent orbitals. The EOM are obtained from the stationary condition of the action functional with respect to any variation of its independent variables, i.e. $\delta S = 0$, or explicitly,
\begin{multline}\label{varia_S}
 \int_{0}^{T}\left[ \langle\delta\Psi(t)| ( i\partial_{t}-H(t)) |\Psi(t)\rangle+\langle\Psi(t)|(-i\overleftarrow{\partial_{t}}-H(t)) |\delta\Psi(t)\rangle \right. \\ 
 \left. + \sum_{\kappa=1}^{K} \sum_{i_{\kappa}j_{\kappa}=1}^{M^{(\kappa)}} \left\{\epsilon_{j_{\kappa}}^{i_{\kappa}(\kappa)}(t)\right(\langle\delta\phi^{(\kappa)}_{i_{\kappa}}(t)|\phi^{(\kappa)}_{j_{\kappa}}(t)\rangle+ \langle\phi^{(\kappa)}_{i_{\kappa}}(t)|\delta\phi^{(\kappa)}_{j_{\kappa}}(t)\rangle\left) \right. \right. \\
 \left. \left. + \delta\epsilon_{j_{\kappa}}^{i_{\kappa}(\kappa)}(t) \left( \langle\phi^{(\kappa)}_{i_{\kappa}}(t)|\phi^{(\kappa)}_{j_{\kappa}}(t)\rangle - \delta_{j_{\kappa}i_{\kappa}}\right)\right\}  \right] = 0,
\end{multline}
where the symbol $\overleftarrow{\partial_{t}}$ denotes that the operator acts to the left. The variation of the wavefunction, $|\delta\Psi(t)\rangle$, reads
\begin{equation}\label{var_Psi}
|\delta\Psi(t)\rangle = \sum_{\vec{n}_{1},\dots,\vec{n}_{K}\in \mathcal{V}_\text{RAS}} \delta C_{\vec{n}_{1},\dots,\vec{n}_{K}}(t)|\vec{n}_{1},\dots,\vec{n}_{K},t\rangle + \sum_{\kappa} \sum_{p_{\kappa}q_{\kappa}}b_{p_{\kappa}}^{\dag(\kappa)}b^{(\kappa)}_{q_{\kappa}}|\Psi(t)\rangle \langle\phi^{(\kappa)}_{p_{\kappa}}(t)|\delta\phi^{(\kappa)}_{q_{\kappa}}(t)\rangle,
\end{equation} 
and its derivative with respect to the time reads
 \begin{equation}\label{time_der_Psi}
\partial_{t}|\Psi(t)\rangle = \sum_{\vec{n}_{1},\dots,\vec{n}_{K}\in \mathcal{V}_\text{RAS}} \dot{C}_{\vec{n}_{1},\dots,\vec{n}_{K}}(t)|\vec{n}_{1},\dots,\vec{n}_{K},t\rangle + \sum_{\kappa}D^{(\kappa)}(t)|\Psi(t)\rangle,
\end{equation} 
with
 \begin{equation}\label{D_kappa}
 D^{(\kappa)}(t) = \sum_{p_{\kappa}q_{\kappa}}b_{p_{\kappa}}^{\dag(\kappa)}b^{(\kappa)}_{q_{\kappa}}\eta_{q_{\kappa}}^{p_{\kappa}(\kappa)}(t).
\end{equation} 
We use $\eta_{q_{\kappa}}^{p_{\kappa}(\kappa)}(t)$ as notation for the matrix element $\eta_{q_{\kappa}}^{p_{\kappa}(\kappa)}(t)=\langle\phi^{(\kappa)}_{p_{\kappa}}(t)|\dot{\phi}^{(\kappa)}_{q_{\kappa}}(t)\rangle$. As detailed below, these $K$ anti-hermitian matrices play an important role in the derivation of the TD-RASSCF EOM. Starting with the variation of the action functional with respect to the Lagrange multipliers and seeking for the stationary condition, we find that $\delta S/\delta\epsilon_{j_{\kappa}^{i_{\kappa}(\kappa)}}=0$ ensures orthonormality of the orbitals at all times, i.e., $\langle\phi^{(\kappa)}_{i_{\kappa}}(t)|\phi^{(\kappa)}_{j_{\kappa}}(t)\rangle=\delta_{i_{\kappa}j_{\kappa}}$, for all $\kappa$.

\subsection{Equations of motion for the amplitudes} 

The stationary condition with respect to a variation of the amplitudes $C_{\vec{n}_{1},\dots,\vec{n}_{K}}^{*}(t)$ in Eq. (\ref{varia_S}) leads to, 
\begin{equation}
\langle\vec{n}_{1},\dots,\vec{n}_{K},t|i\partial_{t}-H(t)|\Psi(t)\rangle = 0 \Leftrightarrow i\dot{C}_{\vec{n}_{1},\dots,\vec{n}_{K}}+\langle\vec{n}_{1},\dots,\vec{n}_{K},t|i \sum_{\kappa}D^{(\kappa)}(t)-H(t)|\Psi(t)\rangle=0, 
\end{equation}
where Eqs. (\ref{var_Psi})-(\ref{time_der_Psi}) were used. Inserting the explicit expressions of the Hamiltonian [Eqs. (\ref{H_tot})-(\ref{H_inter})] and the operators $D^{(\kappa)}(t)$ [Eq. (\ref{D_kappa})] and using the fact that the contributions resulting from the annihilation and creation operators acting on the $\cal{Q}$-space orbitals vanish when the inner product is evaluated, we obtain,
\begin{multline} \label{EOM_C_final}
i \dot{C}_{\vec{n}_{1},\dots,\vec{n}_{K}}(t) =\sum_{\kappa} \left[ \sum_{i_{\kappa}j_{\kappa}} \left(h_{j_{\kappa}}^{i_{\kappa}(\kappa)}(t) - i\eta_{j_{\kappa}}^{i_{\kappa}(\kappa)}(t) \right) \langle\vec{n}_{1},\dots,\vec{n}_{K},t|b_{i_{\kappa}}^{\dag(\kappa)}b^{(\kappa)}_{j_{\kappa}}|\Psi(t)\rangle \right.\\
+\frac{1}{2} \sum_{i_{\kappa}j_{\kappa}k_{\kappa}l_{\kappa}} v_{j_{\kappa}l_{\kappa}}^{i_{\kappa}k_{\kappa}(\kappa)}(t) \langle\vec{n}_{1},\dots,\vec{n}_{K},t|b_{i_{\kappa}}^{\dag(\kappa)}b_{k_{\kappa}}^{\dag(\kappa)}b^{(\kappa)}_{l_{\kappa}}b^{(\kappa)}_{j_{\kappa}}|\Psi(t)\rangle \\ 
\left.+ \sum_{\gamma>\kappa}\sum_{i_{\kappa}i_{\gamma}j_{\kappa}j_{\gamma}}w_{j_{\kappa}j_{\gamma}}^{i_{\kappa}i_{\gamma}(\kappa\gamma)}(t)  \langle\vec{n}_{1},\dots,\vec{n}_{K},t|b_{i_{\kappa}}^{\dag(\kappa)}b^{(\kappa)}_{j_{\kappa}}b_{i_{\gamma}}^{\dag(\gamma)}b^{(\gamma)}_{j_{\gamma}}|\Psi(t)\rangle\right].
\end{multline}
This result is general for any mixtures of bosons and fermions, only the action of the creation and annihilation operators differs. If only particles of a single type are considered, Eq. (\ref{EOM_C_final}) simplifies to the usual TD-RASSCF equation for the amplitudes for bosons or fermions, as obtained in Refs. \cite{Haru13,Haru14_1,Leveque17}. The EOM for the amplitudes, Eq. (\ref{EOM_C_final}), provide an expression for the time derivative of the amplitudes, required for a numerical integration. All quantities that enter Eq. (\ref{EOM_C_final}) can be evaluated solely by the knowledge of the $\mathcal{P}^{(\kappa)}$-space orbitals and the amplitudes at time $t$, except the matrix elements $\eta_{j_{\kappa}}^{i_{\kappa}(\kappa)}(t)$, as discussed in Sec. \ref{EOM_P_orb}.

\subsection{Equations of motion for the orbitals} 
Considering now the EOM for the TD orbitals, the variation of $S$ with respect to an orbital $\langle\phi^{(\kappa)}_{i_{\kappa}}(t)|$ for the particles of type $\kappa$ gives,
\begin{multline}\label{EOM_orb_K}
\frac{\delta S}{\langle\delta\phi^{(\kappa)}_{i_{\kappa}}(t)|}=0 \Leftrightarrow \sum_{j_{\kappa}}\epsilon_{j_{\kappa}}^{i_{\kappa(\kappa)}}(t)|\phi^{(\kappa)}_{j_{\kappa}}(t)\rangle + \sum_{q_{\kappa}} |\phi^{(\kappa)}_{q_{\kappa}}(t)\rangle \langle\Psi(t)|b^{\dag(\kappa)}_{i_{\kappa}}b^{(\kappa)}_{q_{\kappa}} \\
\times \left[\sum_{\vec{n}_{1},\dots,\vec{n}_{K}\in \mathcal{V}_\text{RAS}} i\dot{C}_{\vec{n}_{1},\dots,\vec{n}_{K}}(t)Ê|\vec{n}_{1},\dots,\vec{n}_{K},t\rangle +\left( i\sum_{\kappa}D^{(\kappa)}(t)-H(t)\right)|\Psi(t)\rangle \right] = 0.
\end{multline}
The index $q_{\kappa}$ runs over all the orbitals in the $\mathcal{P}^{(\kappa)}$ and $\mathcal{Q}^{(\kappa)}$ spaces, see Fig. \ref{Fig_1}. The EOM for the orbitals of the $\mathcal{P}^{(\kappa)}$ and $\mathcal{Q}^{(\kappa)}$ spaces, for the particles of type $\kappa$, are obtained by projecting Eq. (\ref{EOM_orb_K}) on an orbital of the $\mathcal{P}^{(\kappa)}$ space, $\langle\phi^{(\kappa)}_{j_{\kappa}}(t)|$ and of the $\mathcal{Q}^{(\kappa)}$ space $\langle\phi^{(\kappa)}_{a_{\kappa}}(t)|$, respectively, and we now consider these two cases individually.   
	
\subsubsection{Equations of motion for the $\cal{Q}$-space orbitals} 

Multiplying Eq. (\ref{EOM_orb_K}) from the left with an orbital  $\langle\phi^{(\kappa)}_{a_{\kappa}}(t)|$ belonging to the $\mathcal{Q}^{(\kappa)}$ space leads to, 
\begin{multline}\label{EOM_orb_Q_K}
 \sum_{\vec{n}_{1},\dots,\vec{n}_{K}\in \cal{V}_\text{RAS}} i\dot{C}_{\vec{n}_{1},\dots,\vec{n}_{K}}(t) \langle\Psi(t)|b_{i_{\kappa}}^{\dag(\kappa)}b^{(\kappa)}_{a_{\kappa}}|\vec{n}_{1},\dots,\vec{n}_{K},t\rangle \\
 +   \langle\Psi(t)|b_{i_{\kappa}}^{\dag(\kappa)}b^{(\kappa)}_{a_{\kappa}} \left( i\sum_{\gamma}D^{(\gamma)}(t)-H(t)\right)|\Psi(t)\rangle= 0,
\end{multline}
where we used the orthonormality between the orbitals of the $\cal{P}^{(\kappa)}$ and $\cal{Q}^{(\kappa)}$ spaces. Equation (\ref{EOM_orb_Q_K}) can be simplified further noting that the inner product, $\langle\Psi(t)|b_{i_{\kappa}}^{\dag(\kappa)}b^{(\kappa)}_{a_{\kappa}}|\vec{n}_{1},\dots,\vec{n}_{K},t\rangle$, vanishes because in all configurations $|\vec{n}_{1},\dots,\vec{n}_{K},t\rangle$ the orbital $|\phi^{(\kappa)}_{a_{\kappa}}(t)\rangle$ is unoccupied. Then, using the explicit expression of the total Hamiltonian [Eqs. (\ref{H_tot}-\ref{H_inter})] and the operators $D^{(\kappa)}(t)$ [Eq. (\ref{D_kappa})] we obtain,
\begin{multline} \label{EOM_with_H_Q_K}
\sum_{p_{\kappa}q_{\kappa}}\left(i\eta_{q_{\kappa}}^{p_{\kappa}(\kappa)}(t) - h_{q_{\kappa}}^{p_{\kappa}(\kappa)}(t)\right)\langle\Psi(t)|b_{i_{\kappa}}^{\dag(\kappa)}b^{(\kappa)}_{a_{\kappa}}b_{p_{\kappa}}^{\dag(\kappa)}b^{(\kappa)}_{q_{\kappa}}|\Psi(t)\rangle\\
+\sum_{\gamma\ne\kappa}\sum_{p_{\gamma}q_{\gamma}}\left(i\eta_{q_{\gamma}}^{p_{\gamma}(\gamma)}(t) - h_{q_{\gamma}}^{p_{\gamma}(\gamma)}(t)\right)\langle\Psi(t)|b_{i_{\kappa}}^{\dag(\kappa)}b^{(\kappa)}_{a_{\kappa}}b_{p_{\gamma}}^{\dag(\gamma)}b^{(\gamma)}_{q_{\gamma}}|\Psi(t)\rangle =  \\ 
\frac{1}{2}\left[\sum_{p_{\kappa}q_{\kappa}r_{\kappa}s_{\kappa}} v_{q_{\kappa}s_{\kappa}}^{p_{\kappa}r_{\kappa} (\kappa)}(t)\langle\Psi(t)|b_{i_{\kappa}}^{\dag(\kappa)}b^{(\kappa)}_{a_{\kappa}}b_{p_{\kappa}}^{\dag(\kappa)}b_{r_{\kappa}}^{\dag(\kappa)}b^{(\kappa)}_{s_{\kappa}}b^{(\kappa)}_{q_{\kappa}}|\Psi(t)\rangle\right.\\
\left.+\sum_{\gamma\ne\kappa}\sum_{p_{\kappa}p_{\gamma}q_{\kappa}q_{\gamma}}w_{q_{\kappa}q_{\gamma}}^{p_{\kappa}p_{\gamma} (\kappa\gamma)}(t)\langle\Psi(t)|b_{i_{\kappa}}^{\dag(\kappa)}b^{(\kappa)}_{a_{\kappa}}b_{p_{\kappa}}^{\dag(\kappa)}b^{(\kappa)}_{q_{\kappa}}b_{p_{\gamma}}^{\dag(\gamma)}b^{(\gamma)}_{q_{\gamma}}|\Psi(t)\rangle \right.\\
\left.+\sum_{\gamma\ne\kappa}\sum_{p_{\gamma}q_{\gamma}r_{\gamma}s_{\gamma}} v_{q_{\gamma}s_{\gamma}}^{p_{\gamma}r_{\gamma} (\gamma)}(t)\langle\Psi(t)|b_{i_{\kappa}}^{\dag(\kappa)}b^{(\kappa)}_{a_{\kappa}}b_{p_{\gamma}}^{\dag(\gamma)}b_{r_{\gamma}}^{\dag(\gamma)}b^{(\gamma)}_{s_{\gamma}}b^{(\gamma)}_{q_{\gamma}}|\Psi(t)\rangle\right.\\
\left. +\sum_{\gamma\ne\kappa}\sum_{\mu\ne\gamma\ne\kappa}\sum_{p_{\gamma}p_{\mu}q_{\gamma}q_{\mu}}w_{q_{\gamma}q_{\mu}}^{p_{\gamma}p_{\mu} (\kappa\mu)}(t)\langle\Psi(t)|b_{i_{\kappa}}^{\dag(\kappa)}b^{(\kappa)}_{a_{\kappa}}b_{p_{\gamma}}^{\dag(\gamma)}b^{(\gamma)}_{q_{\gamma}}b_{p_{\mu}}^{\dag(\mu)}b^{(\mu)}_{q_{\mu}}|\Psi(t)\rangle\right].
\end{multline}  
Using the (anti-)commutation relations for the creation and annihilation operators for (fermions) bosons, we now reestablish the normal ordering of the chains of operators in Eq. (\ref{EOM_with_H_Q_K}). The case of a chain of operators acting on a single type of particles is discussed in detail in Ref. \cite{Leveque17} and will not be addressed here. Since the operators acting on different types of particles commute, the chains including such operators can be expressed as,
\begin{eqnarray}
&&b_{i_{\kappa}}^{\dag(\kappa)}b^{(\kappa)}_{a_{\kappa}}b_{p_{\gamma}}^{\dag(\gamma)}b^{(\gamma)}_{q_{\gamma}}=b_{p_{\gamma}}^{\dag(\gamma)}b^{(\gamma)}_{q_{\gamma}} b_{i_{\kappa}}^{\dag(\kappa)}b^{(\kappa)}_{a_{\kappa}}\label{four_op_1}, \\ 
&&b_{i_{\kappa}}^{\dag(\kappa)}b^{(\kappa)}_{a_{\kappa}}b_{p_{\gamma}}^{\dag(\gamma)}b_{r_{\gamma}}^{\dag(\gamma)}b^{(\gamma)}_{s_{\gamma}}b^{(\gamma)}_{q_{\gamma}}=b_{p_{\gamma}}^{\dag(\gamma)}b_{r_{\gamma}}^{\dag(\gamma)}b^{(\gamma)}_{s_{\gamma}}b^{(\gamma)}_{q_{\gamma}}b_{i_{\kappa}}^{\dag(\kappa)}b^{(\kappa)}_{a_{\kappa}} \label{six_op_1},  \\
&&b_{i_{\kappa}}^{\dag(\kappa)}b^{(\kappa)}_{a_{\kappa}}b_{p_{\gamma}}^{\dag(\gamma)}b^{(\gamma)}_{q_{\gamma}}b_{p_{\mu}}^{\dag(\mu)}b^{(\mu)}_{q_{\mu}}=b_{p_{\gamma}}^{\dag(\gamma)}b^{(\gamma)}_{q_{\gamma}}b_{p_{\mu}}^{\dag(\mu)}b^{(\mu)}_{q_{\mu}}b_{i_{\kappa}}^{\dag(\kappa)}b^{(\kappa)}_{a_{\kappa}}  \label{six_op_2},  \\
&&b_{i_{\kappa}}^{\dag(\kappa)}b^{(\kappa)}_{a_{\kappa}}b_{p_{\kappa}}^{\dag(\kappa)}b^{(\kappa)}_{q_{\kappa}}b_{p_{\gamma}}^{\dag(\gamma)}b^{(\gamma)}_{q_{\gamma}}=b_{p_{\gamma}}^{\dag(\gamma)}b^{(\gamma)}_{q_{\gamma}}b_{i_{\kappa}}^{\dag(\kappa)}b^{(\kappa)}_{q_{\kappa}}\delta_{a_{\kappa}p_{\kappa}}+b_{p_{\gamma}}^{\dag(\gamma)}b^{(\gamma)}_{q_{\gamma}}b_{i_{\kappa}}^{\dag(\kappa)}b_{p_{\kappa}}^{\dag(\kappa)}b^{\kappa}_{q_{\kappa}}b^{(\kappa)}_{a_{\kappa}}.  \label{six_op_3}
\end{eqnarray} 
In the case of Eqs. (\ref{four_op_1}-\ref{six_op_2}) the chains of operators create a particle $\kappa$ in the $\mathcal{Q}^{(\kappa)}$-space orbital $|\phi^{(\kappa)}_{a_{\kappa}}(t)\rangle$, which is always unoccupied in the total wavefunction, thus the inner product including these terms vanishes. The same result holds for the second term of Eq. (\ref{six_op_3}), thus only the first one with four operators remains. The summations in Eq. (\ref{EOM_with_H_Q_K}) can now be restricted to the $\mathcal{P}^{(\kappa)}$-space orbitals as the contribution arising from the $\mathcal{Q}^{(\kappa)}$-space orbitals vanishes. Using the symmetry properties of the matrix elements, $v_{j_{\kappa}l_{\kappa}}^{a_{\kappa}k_{\kappa}(\kappa)}(t)=v_{l_{\kappa}j_{\kappa}}^{k_{\kappa}a_{\kappa}(\kappa)}(t)$, Eq. (\ref{EOM_with_H_Q_K}) simplifies to,
\begin{equation}\label{EOM_with_H_Q_K_rho}
 \sum_{j_{\kappa}}\left( i\eta_{j_{\kappa}}^{a_{\kappa}(\kappa)}(t)-h_{j_{\kappa}}^{a_{\kappa}(\kappa)}(t) \right) \rho_{i_{\kappa}}^{j_{\kappa}(\kappa)}(t) =\sum_{j_{\kappa}k_{\kappa}l_{\kappa}}v^{a_{\kappa}k_{\kappa}(\kappa)}_{j_{\kappa}l_{\kappa}}(t) \rho_{i_{\kappa}k_{\kappa}}^{j_{\kappa}l_{\kappa}(\kappa)}(t)+\frac{1}{2}\sum_{\gamma\ne\kappa}\sum_{j_{\gamma}k_{\kappa}k_{\gamma}}w^{a_{\kappa}j_{\gamma}(\kappa\gamma)}_{k_{\kappa}k_{\gamma}}(t) \rho_{i_{\kappa}j_{\gamma}}^{k_{\kappa}k_{\gamma}(\kappa\gamma)}(t),
\end{equation}
where the reduced one- and two-body density matrices are defined by, 
\begin{eqnarray}
 \rho_{i_{\kappa}}^{j_{\kappa}(\kappa)}(t) &=& \langle \Psi(t)|b_{i_{\kappa}}^{\dag(\kappa)}b^{(\kappa)}_{j_{\kappa}}|\Psi(t) \rangle,\\
 \rho_{i_{\kappa}k_{\kappa}}^{j_{\kappa}l_{\kappa}(\kappa)}(t) &=& \langle \Psi(t)|b_{i_{\kappa}}^{\dag(\kappa)}b_{k_{\kappa}}^{\dag(\kappa)}b^{(\kappa)}_{l_{\kappa}}b^{(\kappa)}_{j_{\kappa}}|\Psi(t) \rangle,\\
 \rho_{i_{\kappa}j_{\gamma}}^{k_{\kappa}k_{\gamma}(\kappa\gamma)}(t) &=& \langle \Psi(t)|b_{i_{\kappa}}^{\dag(\kappa)}b^{(\kappa)}_{k_{\kappa}}b_{j_{\gamma}}^{\dag(\gamma)}b^{(\gamma)}_{k_{\gamma}}|\Psi(t) \rangle. \label{rho_mix}
\end{eqnarray}
Equation (\ref{EOM_with_H_Q_K_rho}) contains the explicit consideration of the $\cal{Q}^{(\kappa)}$-space orbital $|\phi^{(\kappa)}_{a_{\kappa}}(t)\rangle$. There is an infinite number of virtual orbitals in the $\cal{Q}^{(\kappa)}$ space and their explicit consideration is therefore not possible. Thus, to circumvent the explicit consideration of these unoccupied orbitals, the projector onto the space spanned by the $\cal{Q}^{(\kappa)}$-space orbitals $Q^{(\kappa)}(t)=\sum_{a_{\kappa}}|\phi^{(\kappa)}_{a_{\kappa}}(t)\rangle\langle\phi^{(\kappa)}_{a_{\kappa}}(t)|$ is used to consider the role of the $\cal{Q}^{(\kappa)}$-space orbitals implicitly. The EOM for the $\cal{Q}^{(\kappa)}$-space orbitals finally reads,
\begin{multline}\label{final_eq_Q_K}
 i\sum_{j_{\kappa}} Q^{(\kappa)}(t)|\dot{\phi}^{(\kappa)}_{j_{\kappa}}(t)\rangle \rho_{i_{\kappa}}^{j_{\kappa}(\kappa)}(t) = 
 Q^{(\kappa)}(t)\left[
 \sum_{j_{\kappa}} h^{(\kappa)}(\textbf{x}_{\kappa},t)|\phi^{(\kappa)}_{j_{\kappa}}(t)\rangle \rho_{i_{\kappa}}^{j_{\kappa}(\kappa)}(t)Ê\right.\\
 \left. + \sum_{j_{\kappa}k_{\kappa}l_{\kappa}} V_{l_{\kappa}}^{k_{\kappa}(\kappa)}(\textbf{x}_{\kappa},t)|\phi^{(\kappa)}_{j_{\kappa}}(t)\rangle \rho_{i_{\kappa}k_{\kappa}}^{j_{\kappa}l_{\kappa}(\kappa)}(t)
  + \frac{1}{2}\sum_{\gamma\ne\kappa}\sum_{j_{\gamma}j_{\kappa}k_{\gamma}}W^{j_{\gamma}(\kappa\gamma)}_{k_{\gamma}}(\textbf{x}_{\kappa},t)|\phi^{(\kappa)}_{j_{\kappa}}(t)\rangle \rho_{i_{\kappa}j_{\gamma}}^{j_{\kappa}k_{\gamma}(\kappa\gamma)} (t)
 \right],
\end{multline}
with $V_{l_{\kappa}}^{k_{\kappa}(\kappa)}(\textbf{x}_{\kappa},t)$ and $W^{j_{\gamma}(\kappa\gamma)}_{k_{\gamma}}(\textbf{x}_{\kappa},t)$ the mean-field operators defined as,
\begin{eqnarray}
V_{l_{\kappa}}^{k_{\kappa}(\kappa)}(\textbf{x}_{\kappa},t) &=&Ê\int{\phi_{k_{\kappa}}^{*(\kappa)}(\textbf{x}'_{\kappa},t)}v^{(\kappa)}(\textbf{x}_{\kappa},\textbf{x}'_{\kappa},t)\phi^{(\kappa)}_{l_{\kappa}}(\textbf{x}'_{\kappa},t)d\textbf{x}'_{\kappa}, \label{MF_inter} \\ 
W^{j_{\gamma}(\kappa\gamma)}_{k_{\gamma}}(\textbf{x}_{\kappa},t)&=&Ê\int{\phi^{*(\gamma)}_{j_{\gamma}}(\textbf{x}_{\gamma},t)}w^{(\kappa\gamma)}(\textbf{x}_{\kappa},\textbf{x}_{\gamma},t)\phi^{(\gamma)}_{k_{\gamma}}(\textbf{x}_{\gamma},t)d\textbf{x}_{\gamma}. \label{MF_intra}
\end{eqnarray}
The mean-field of Eq. (\ref{MF_inter}) results from the interaction between the particles of type $\kappa$ in the system, while Eq. (\ref{MF_intra}) describes the interaction between the particles of type $\kappa$ and the other types of particles. The $\mathcal{Q}$-space EOM can be evaluated as all quantities that enter Eq. (\ref{final_eq_Q_K}) are known at a time $t$. 
Solving the $\mathcal{Q}$-space orbital equations, provides one of the components of the time derivative of the $\mathcal{P}$-space orbitals, namely, $Q^{(\kappa)}(t)|\dot{\phi}^{(\kappa)}_{i_{\kappa}}(t)\rangle$, see Eq. (\ref{deriv_orb}).

\subsubsection{Equations of motion for the $\cal{P}$-space orbitals}\label{EOM_P_orb} 

To obtain the EOM for the $\mathcal{P}$-space orbitals, we multiply Eq. (\ref{EOM_orb_K}) on the left by an orbital of the $\cal{P}^{(\kappa)}$ space, $\langle\phi^{(\kappa)}_{j_{\kappa}}(t)|$, and obtain
\begin{multline}\label{EOM_orb_P_1_K}
\sum_{\vec{n}_{1},\dots,\vec{n}_{K}\in \cal{V}_\text{RAS}} i\dot{C}_{\vec{n}_{1},\dots,\vec{n}_{K}}(t) \langle\Psi(t)|b_{i_{\kappa}}^{\dag(\kappa)}b_{j_{\kappa}}^{(\kappa)}Ê|\vec{n}_{1},\dots,\vec{n}_{K},t\rangle \\
+  \langle\Psi(t)|b_{i_{\kappa}}^{\dag(\kappa)}b_{j_{\kappa}}^{(\kappa)} \left( i\sum_{\gamma}D^{(\gamma)}(t)-H(t)\right)|\Psi(t)\rangle + \epsilon_{j_{\kappa}}^{i_{\kappa}(\kappa)}(t) = 0.
\end{multline}
This equation still contains the Lagrange multiplier $ \epsilon_{j_{\kappa}}^{i_{\kappa}(\kappa)}(t)$. A variation of $S$ with respect to the orbital $|\phi^{(\kappa)}_{j_{\kappa}}(t)\rangle$ and its projection onto an orbital $\langle\phi^{(\kappa)}_{i_{\kappa}}(t)|$, leads to an equation containing the same Lagrange multiplier, 
\begin{multline}\label{EOM_orb_P_2_K}
\sum_{\vec{n}_{1},\dots,\vec{n}_{K}\in \cal{V}_\text{RAS}} -i\dot{C}^{*}_{\vec{n}_{1}\dots,\vec{n}_{K}}(t) \langleÊ\vec{n}_{1},\dots,\vec{n}_{K},t|b_{i_{\kappa}}^{\dag(\kappa)}b_{j_{\kappa}}^{(\kappa)}|\Psi(t)\rangle \\
+  \langle\Psi(t)| \left( i\sum_{\gamma}D^{(\gamma)}(t)-H(t)\right)b_{i_{\kappa}}^{\dag(\kappa)}b_{j_{\kappa}}^{(\kappa)}|\Psi(t)Ê\rangle + \epsilon_{j_{\kappa}}^{i_{\kappa}(\kappa)}(t) = 0.
\end{multline}
Subtracting Eqs. (\ref{EOM_orb_P_1_K}) and (\ref{EOM_orb_P_2_K}) removes the Lagrange multipliers,
 \begin{multline} \label{P_space_general_K}
i\dot{\rho}_{i_{\kappa}}^{j_{\kappa}(\kappa)}(t)=\\
\langle\Psi(t)| \left( i\sum_{\gamma}D^{(\gamma)}(t)-H(t)\right)b_{i_{\kappa}}^{\dag(\kappa)}b_{j_{\kappa}}^{(\kappa)}|\Psi(t)Ê\rangle - \langle\Psi(t)|b_{i_{\kappa}}^{\dag(\kappa)}b_{j_{\kappa}}^{(\kappa)} \left( i\sum_{\gamma}D^{(\gamma)}(t)-H(t)\right)|\Psi(t)\rangle.
\end{multline}      
Here we have introduced 
\begin{equation}
\dot{\rho}_{i_{\kappa}}^{j_{\kappa}(\kappa)}(t)\equiv\sum_{\vec{n}_{1},\dots,\vec{n}_{K}\in \cal{V}_\text{RAS}}(\dot{C}^{*}_{\vec{n}_{1},\dots,\vec{n}_{K}}(t)\langleÊ\vec{n}_{1},\dots,\vec{n}_{K},t|b_{i_{\kappa}}^{\dag(\kappa)}b_{j_{\kappa}}^{(\kappa)}|\Psi(t)\rangle+\langle\Psi(t)|b_{i_{\kappa}}^{\dag(\kappa)}b_{j_{\kappa}}^{(\kappa)}Ê|\vec{n}_{1},\dots,\vec{n}_{K},t\rangle\dot{C}_{\vec{n}_{1},\dots,\vec{n}_{K}}(t)),
\end{equation} 
the time derivative of the reduced one-body density matrix. From Eq. (\ref{P_space_general_K}), we can collect terms including the operator $D^{(\gamma)}(t)$ with $\gamma\ne\kappa$,
 \begin{multline}
i\dot{\rho}_{i_{\kappa}}^{j_{\kappa}(\kappa)}(t)=
\langle\Psi(t)| i\sum_{\gamma\ne\kappa}D^{(\gamma)}(t)b_{i_{\kappa}}^{\dag(\kappa)}b_{j_{\kappa}}^{(\kappa)}|\Psi(t)Ê\rangle - \langle\Psi(t)|b_{i_{\kappa}}^{\dag(\kappa)}b_{j_{\kappa}}^{(\kappa)}  i\sum_{\gamma\ne\kappa}D^{(\gamma)}(t)|\Psi(t)\rangle\\
\langle\Psi(t)| \left( iD^{(\kappa)}(t)-H(t)\right)b_{i_{\kappa}}^{\dag(\kappa)}b_{j_{\kappa}}^{(\kappa)}|\Psi(t)Ê\rangle - \langle\Psi(t)|b_{i_{\kappa}}^{\dag(\kappa)}b_{j_{\kappa}}^{(\kappa)} \left( iD^{(\kappa)}(t)-H(t)\right)|\Psi(t)\rangle.
\end{multline}   
Using the explicit expression of $D^{(\gamma)}(t)$ from Eq. (\ref{D_kappa}), we can show that for $\gamma\ne\kappa$, 
\begin{multline} \label{simplify_P_eq}
\sum_{\gamma\ne\kappa}\sum_{p_{\gamma}q_{\gamma}}\eta_{q_{\gamma}}^{p_{\gamma}(\gamma)}(t)\langle\Psi(t)|b_{p_{\gamma}}^{\dag(\gamma)}b^{(\gamma)}_{q_{\gamma}}b_{i_{\kappa}}^{\dag(\kappa)}b_{j_{\kappa}}^{(\kappa)}|\Psi(t)\rangle - \sum_{\gamma\ne\kappa}\sum_{p_{\gamma}q_{\gamma}}\eta_{q_{\gamma}}^{p_{\gamma}(\gamma)}(t)\langle\Psi(t)|b_{i_{\kappa}}^{\dag(\kappa)}b_{j_{\kappa}}^{(\kappa)}b_{p_{\gamma}}^{\dag(\gamma)}b^{(\gamma)}_{q_{\gamma}}|\Psi(t)\rangle \\
= \sum_{\gamma\ne\kappa}\sum_{p_{\gamma}q_{\gamma}}\eta_{q_{\gamma}}^{p_{\gamma}(\gamma)}(t)\langle\Psi(t)|b_{p_{\gamma}}^{\dag(\gamma)}b^{(\gamma)}_{q_{\gamma}}b_{i_{\kappa}}^{\dag(\kappa)}b_{j_{\kappa}}^{(\kappa)}|\Psi(t)\rangle -\sum_{\gamma\ne\kappa}\sum_{p_{\gamma}q_{\gamma}}\eta_{q_{\gamma}}^{p_{\gamma}(\gamma)}(t)\langle\Psi(t)|b_{p_{\gamma}}^{\dag(\gamma)}b^{(\gamma)}_{q_{\gamma}}b_{i_{\kappa}}^{\dag(\kappa)}b_{j_{\kappa}}^{(\kappa)}|\Psi(t)\rangle   \\
=0.
\end{multline} 
Equation (\ref{simplify_P_eq}) shows that the $\gamma\ne\kappa$ terms do not contribute in the sums over $\gamma$ in Eq.  (\ref{P_space_general_K}). Hence, Eq.  (\ref{P_space_general_K}) simplifies to 
 \begin{equation} \label{P_space_general_K_final}
 i\dot{\rho}_{i_{\kappa}}^{j_{\kappa}(\kappa)}(t)=\langle\Psi(t)| \left( iD^{(\kappa)}(t)-H(t)\right)b_{i_{\kappa}}^{\dag(\kappa)}b_{j_{\kappa}}^{(\kappa)}|\Psi(t)Ê\rangle - \langle\Psi(t)|b_{i_{\kappa}}^{\dag(\kappa)}b_{j_{\kappa}}^{(\kappa)} \left( iD^{(\kappa)}(t)-H(t)\right)|\Psi(t)\rangle.
\end{equation}   
Note that taking into consideration only particles of a single type, this equation simplifies to the one obtained in Refs. \cite{Haru13,Haru14_1,Leveque17}. 

Equation (\ref{P_space_general_K_final}) needs to be solved to obtain 
$\eta_{i_{\kappa}}^{j_{\kappa}(\kappa)}(t)$ necessary for the evaluation of 
the $\cal{P}^{(\kappa)}$-space component of the time-derivative of the $\cal{P}^{(\kappa)}$-space orbitals, see Eq. (\ref{deriv_orb}),  
\begin{equation}\label{time_deriv_Porb}
P^{(\kappa)}(t)|\dot{\phi}^{(\kappa)}_{i_{\kappa}}(t)\rangle =  \sum_{j_{\kappa}}|\dot{\phi}^{(\kappa)}_{j_{\kappa}}(t)\rangle\eta_{i_{\kappa}}^{j_{\kappa}(\kappa)}(t).
\end{equation}
The equations (\ref{P_space_general_K_final}) are coupled to the EOM for the amplitudes, Eq. (\ref{EOM_C_final}), through $\dot{\rho}_{i_{\kappa}}^{j_{\kappa}(\kappa)}(t)$. It is important to note that these equations are not coupled directly to each other but only trough the time-derivative of the reduced density matrices. This issue can be easily dealt with when there are no ristrictions on the active orbital space as in the MCTDHF, MCTDHB and MCTDH for mixtures because the wavefunction is invariant under a simultaneous unitary transformation of the orbitals and its reverse applied to the amplitudes. This provides the gauge freedom to choose $\eta_{i_{\kappa}}^{j_{\kappa}(\kappa)}(t)=0$ for the sets of orbitals for each $\kappa$. In the case of the wavefunction based on the RAS \textit{Ansatz}, such a freedom in the choice of the matrix elements $\eta_{i_{\kappa}}^{j_{\kappa}(\kappa)}(t)$ is not possible, except for pairs of orbitals which belong to the same $P^{(\kappa)}_{i_{\kappa}}$ ($i_{\kappa}=0,1,2$) space. Thus, Eq. (\ref{P_space_general_K_final}) must be solved for pairs of orbitals $\{i_{\kappa}',j_{\kappa}''\}$, which belong to different $P^{(\kappa)}_{i_{\kappa}}$ space. In the following, we use the prime ($'$) and double prime ($''$) symbols to indicate that the orbitals belong to two different $P^{(\kappa)}_{i_{\kappa}}$ spaces. 
\\

\paragraph{Even excitation RAS scheme} \hspace{0pt}  \label{Even excitation RAS scheme} \\

 In the derivation of the TD-RASSCF EOM for electrons \cite{Haru13,Haru14_1}, two methods were proposed to circumvent  the coupling of the amplitudes and $\cal{P}$-space orbitals in their EOM. The first one is to consider only \textit{even} excitations of particles between the $\mathcal{P}_{1}$ and $\mathcal{P}_{2}$ orbitals. In the case of a mixture of different types of particles, we consider this restriction for transitions from  $\mathcal{P}^{(\kappa)}_{1}$ to $\mathcal{P}^{(\kappa)}_{2}$. Writing explicitly $\dot{\rho}_{i'_{\kappa}}^{j''_{\kappa}(\kappa)}$, 
\begin{multline}\label{TD_rho_K} 
\dot{\rho}_{i'_{\kappa}}^{j''_{\kappa}(\kappa)}(t)=\\
\sum_{\vec{n}_{1},\dots,\vec{n}_{K}\in \cal{V}_\text{RAS}}(\dot{C}^{*}_{\vec{n}_{1},\dots,\vec{n}_{K}}(t)\langleÊ\vec{n}_{1},\dots,\vec{n}_{K},t|b_{i'_{\kappa}}^{\dag(\kappa)}b_{j''_{\kappa}}^{(\kappa)}|\Psi(t)\rangle+\langle\Psi(t)|b_{i'_{\kappa}}^{\dag(\kappa)}b_{j''_{\kappa}}^{(\kappa)}Ê|\vec{n}_{1},\dots,\vec{n}_{K},t\rangle\dot{C}_{\vec{n}_{1},\dots,\vec{n}_{K}}(t)), 
\end{multline} 
the action of $b_{i'_{\kappa}}^{\dag(\kappa)}b^{(\kappa)}_{j''_{\kappa}}$ on $|\Psi(t)\rangle$ annihilates one particle of type $\kappa$ in the $\mathcal{P}^{(\kappa)}_{2}$ orbital $|\phi^{(\kappa)}_{j''_{\kappa}}(t)\rangle$ and creates one in the $\mathcal{P}^{(\kappa)}_{1}$ orbital $|\phi^{(\kappa)}_{i'_{\kappa}}(t)\rangle$. Since only even numbers of particles are present in the $\mathcal{P}^{(\kappa)}_{2}$ orbitals in the even excitation RAS scheme, it follows that $b_{i'_{\kappa}}^{\dag(\kappa)}b^{(\kappa)}_{j''_{\kappa}}|\Psi(t)\rangle$ contains only configurations with an odd number of particles in  
$\mathcal{P}^{(\kappa)}_{2}$. Thus, the inner product with $\langle\vec{n}_{1},\dots,\vec{n}_{K},t|, \forall \{\vec{n}_{1},\dots,\vec{n}_{K}\} \in \{ \mathcal{V}_{\text{RAS}_{1}},\dots,\mathcal{V}_{\text{RAS}_{K}}\}$ is zero. The same result is also obtained when $b_{i'_{\kappa}}^{\dag(\kappa)}b^{(\kappa)}_{j''_{\kappa}}$ acts on the configuration $|\vec{n}_{1},\dots,\vec{n}_{K},t\rangle$ and the inner product with $\langle\Psi(t)|$ is evaluated. It turns out, as obtained in the case of a single type of particle \cite{Haru13,Haru14_1,Leveque17}, that in this specific excitation scheme $\dot{\rho}_{i'_{\kappa}}^{j''_{\kappa}(\kappa)}(t)=0$, for all pairs of orbitals $\{i'_{\kappa},j''_{\kappa}\}$ for each $\kappa$. Because the amplitudes [Eq. (\ref{EOM_C_final})] and the $\mathcal{P}^{(\kappa)}$-space orbital EOM [Eq. (\ref{P_space_general_K_final})] are only coupled through the time-derivative of the reduced one-body density matrix, in the even excitations RAS scheme, the EOM are uncoupled.

The explicit formulation of the EOM for the particle of type $\kappa$ is obtained by inserting the expression of the Hamiltonian [Eqs. (\ref{H_tot})] into Eq. (\ref{P_space_general_K_final}) with $\dot{\rho}_{i'_{\kappa}}^{j''_{\kappa}(\kappa)}(t)=0$.
\begin{multline}\label{eta_eq_even_intermed_K_1}
\langle\Psi(t)| \left( iD^{(\kappa)}(t) - H_{\kappa}(t) - \frac{1}{2}\sum_{\gamma\ne\kappa} H_{\kappa\gamma}(t)\right)b_{i'_{\kappa}}^{\dag(\kappa)}b_{j''_{\kappa}}^{(\kappa)}|\Psi(t)Ê\rangle \\
- \langle\Psi(t)|b_{i'_{\kappa}}^{\dag(\kappa)}b_{j''_{\kappa}}^{(\kappa)} \left( iD^{(\kappa)}(t)- H_{\kappa}(t) - \frac{1}{2}\sum_{\gamma\ne\kappa} H_{\kappa\gamma}(t)\right)|\Psi(t)\rangle\\
-\sum_{\gamma\ne\kappa}\langle\Psi(t)| \left( H_{\gamma}(t) + \frac{1}{2}\sum_{\mu\ne\gamma\ne\kappa} H_{\gamma\mu}(t)\right)b_{i'_{\kappa}}^{\dag(\kappa)}b_{j''_{\kappa}}^{(\kappa)}|\Psi(t)Ê\rangle \\
+\sum_{\gamma\ne\kappa}\langle\Psi(t)|b_{i'_{\kappa}}^{\dag(\kappa)}b_{j''_{\kappa}}^{(\kappa)} \left( H_{\gamma}(t) + \frac{1}{2}\sum_{\mu\ne\gamma\ne\kappa} H_{\gamma\mu}(t)\right)|\Psi(t)Ê\rangle =0.
\end{multline}  
 The Hamiltonians $H_{\gamma}(t)$ and $H_{\gamma\mu}(t)$ with $\{\gamma,\mu\}\ne\kappa$ include creation and annihilation operators that do not act on the particle of type $\kappa$, hence these Hamiltonians commute with $b_{i'_{\kappa}}^{\dag(\kappa)}b^{(\kappa)}_{j''_{\kappa}}$ and the two last terms of Eq. (\ref{eta_eq_even_intermed_K_1}) cancel each other. We obtain, 
\begin{equation}\label{eta_eq_even_intermed_K}
\langle\Psi (t)|\left[b_{i'_{\kappa}}^{\dag(\kappa)}b^{(\kappa)}_{j''_{\kappa}},\left( iD^{(\kappa)}(t) - H_{\kappa}(t) - \frac{1}{2}\sum_{\gamma\ne\kappa} H_{\kappa\gamma}(t)\right)\right]\Psi(t)\rangle =0.
\end{equation}   
This equation can be further simplified using the explicit expressions for $H_{\kappa}(t)$ [Eq. (\ref{H_intra})], $H_{\kappa\gamma}(t)$ [Eq. (\ref{H_inter})] and $D^{(\kappa)}(t)$ [Eq. (\ref{D_kappa})]. The two terms, $H_{\kappa}(t)$ and $D^{(\kappa)}(t)$, in Eq. (\ref{eta_eq_even_intermed_K}) include only operators acting on the particles of type $\kappa$. They are identical to the ones obtained in the case of the TD-RASSCF theory for a single type of fermion \cite{Haru13,Haru14_1} or boson \cite{Leveque17}, and will not be discussed here. The last term, $-\frac{1}{2}\sum_{\gamma\ne\kappa} H_{\kappa\gamma}(t)$, results from the interaction between different types of particles. We explicitly have
\begin{multline}\label{intra_inter}
\langle \Psi(t)|[b_{i'_{\kappa}}^{\dag(\kappa)}b^{(\kappa)}_{j''_{\kappa}},H_{\kappa\gamma}(t)]|\Psi(t)\rangle =\\
 \sum_{p_{\kappa}p_{\gamma}q_{\kappa}q_{\gamma}}w_{q_{\kappa}q_{\gamma}}^{p_{\kappa}p_{\gamma}(\kappa\gamma)}\langle \Psi(t)|[b_{i'_{\kappa}}^{\dag(\kappa)}b^{(\kappa)}_{j''_{\kappa}}b_{p_{\kappa}}^{\dag(\kappa)}b^{(\kappa)}_{q_{\kappa}}b_{p_{\gamma}}^{\dag(\gamma)}b^{(\gamma)}_{q_{\gamma}}
 -b_{p_{\kappa}}^{\dag(\kappa)}b^{(\kappa)}_{q_{\kappa}}b_{p_{\gamma}}^{\dag(\gamma)}b^{(\gamma)}_{q_{\gamma}}b_{i'_{\kappa}}^{\dag(\kappa)}b^{(\kappa)}_{j''_{\kappa}}|\Psi(t)\rangle.
\end{multline}
The two chains of operators on the rhs of Eq. (\ref{intra_inter}) can be written in the normal ordering as,
\begin{eqnarray}
b_{i'_{\kappa}}^{\dag(\kappa)}b^{(\kappa)}_{j''_{\kappa}}b_{p_{\kappa}}^{\dag(\kappa)}b^{(\kappa)}_{q_{\kappa}}b_{p_{\gamma}}^{\dag(\gamma)}b^{(\gamma)}_{q_{\gamma}} =b_{i'_{\kappa}}^{\dag(\kappa)}b^{(\kappa)}_{q_{\kappa}}b_{p_{\gamma}}^{\dag(\gamma)}b^{(\gamma)}_{q_{\gamma}}\delta_{j''_{\kappa}p_{\kappa}}\pm b_{p_{\kappa}}^{\dag(\kappa)}b_{i'_{\kappa}}^{\dag(\kappa)}b^{(\kappa)}_{q_{\kappa}}b^{(\kappa)}_{j''_{\kappa}}b_{p_{\gamma}}^{\dag(\gamma)}b^{(\gamma)}_{q_{\gamma}},\label{chain_simp_1} \\
b_{p_{\kappa}}^{\dag(\kappa)}b^{(\kappa)}_{q_{\kappa}}b_{p_{\gamma}}^{\dag(\gamma)}b^{(\gamma)}_{q_{\gamma}}b_{i'_{\kappa}}^{\dag(\kappa)}b^{(\kappa)}_{j''_{\kappa}}=b_{p_{\kappa}}^{\dag(\kappa)}b^{(\kappa)}_{j''_{\kappa}}b_{p_{\gamma}}^{\dag(\gamma)}b^{(\gamma)}_{q_{\gamma}}\delta_{q_{\kappa}i'_{\kappa}}\pm b_{p_{\kappa}}^{\dag(\kappa)}b_{i'_{\kappa}}^{\dag(\kappa)}b^{(\kappa)}_{q_{\kappa}}b^{(\kappa)}_{j''_{\kappa}}b_{p_{\gamma}}^{\dag(\gamma)}b^{(\gamma)}_{q_{\gamma}},\label{chain_simp_2}
\end{eqnarray}
such that only two chains of four operators remain, while the chains of six operators cancel each other when subtracting Eq. (\ref{chain_simp_2}) from Eq. (\ref{chain_simp_1}) as needed for the rhs of Eq. (\ref{intra_inter}). Finally, the simplified EOM for the particles of type $\kappa$ reads, 
\begin{multline}\label{final_P_space_even_K}
\sum_{k''_{\kappa}l'_{\kappa}}\left(h_{l'_{\kappa}}^{k''_{\kappa}(\kappa)}(t)-i\eta_{l'_{\kappa}}^{k''_{\kappa}(\kappa)}(t)\right)A_{k''_{\kappa}i'_{\kappa}}^{l'_{\kappa}j''_{\kappa}(\kappa)}(t)+\sum_{k_{\kappa}l_{\kappa}n_{\kappa}}(v_{k_{\kappa}l_{\kappa}}^{j''_{\kappa}n_{\kappa}(\kappa)}(t)\rho_{i'_{\kappa}n_{\kappa}}^{k_{\kappa}l_{\kappa}(\kappa)}(t)-v_{i'_{\kappa}n_{\kappa}}^{k_{\kappa}l_{\kappa}(\kappa)}(t)\rho_{k_{\kappa}l_{\kappa}}^{j''_{\kappa}n_{\kappa}(\kappa)}(t)) \\
+\frac{1}{2}\sum_{\gamma\ne\kappa}\sum_{k_{\gamma}k_{\kappa}l_{\gamma}}(w_{k_{\kappa}l_{\gamma}}^{j''_{\kappa}k_{\gamma}(\kappa\gamma)}(t)\rho_{i'_{\kappa}k_{\gamma}}^{k_{\kappa}l_{\gamma}(\kappa\gamma)}(t)-w_{i'_{\kappa}l_{\gamma}}^{k_{\kappa}k_{\gamma}(\kappa\gamma)}(t)\rho_{k_{\kappa}k_{\gamma}}^{j''_{\kappa}l_{\gamma}(\kappa\gamma)}(t)) =0,
\end{multline}
with $A_{k''_{\kappa}i'_{\kappa}}^{l'_{\kappa}j''_{\kappa}(\kappa)}(t)=\rho_{i'_{\kappa}}^{l'_{\kappa}(\kappa)}\delta_{j''_{\kappa}k''_{\kappa}}-\rho_{k''_{\kappa}}^{j''_{\kappa}(\kappa)}\delta_{l'_{\kappa}i'_{\kappa}}$. The EOM for the $\cal{P}$-space orbitals using the even excitation scheme, Eq. (\ref{final_P_space_even_K}), can be solved to obtain the matrix elements $\eta_{i'_{\kappa}}^{j''_{\kappa}(\kappa)}(t)$ for orbitals which belong to different $\mathcal{P}^{(\kappa)}$-spaces for the particles of type $\kappa$. Note that the matrix $\eta^{(\kappa)}(t)$ is anti-hermitian, i.e. $\eta_{i_{\kappa}}^{j_{\kappa}(\kappa)}(t)=-(\eta_{j_{\kappa}}^{i_{\kappa}(\kappa)}(t))^{*}$, thus only the upper (or lower) off-diagonal elements need be determined. Using the even excitation scheme, we obtain one of the components of the time derivative of the $\mathcal{P}^{(\kappa)}$-space orbitals, namely $P^{(\kappa)}(t)|\dot{\phi}^{(\kappa)}_{i_{\kappa}}(t)\rangle=\sum^{M^{(\kappa)}}_{j_{\kappa}}\eta^{j_{\kappa}}_{i_{\kappa}}(t)|\phi^{(\kappa)}_{j_{\kappa}}(t)\rangle$, see Eqs. (\ref{deriv_orb}) and (\ref{time_deriv_Porb}). Combined with the EOM for the $\cal{Q}$-space orbitals, Eq. (\ref{final_eq_Q_K}), the time derivative of the $\mathcal{P}^{(\kappa)}$-space orbitals can be evaluated. In addition, the amplitude equation, Eq. (\ref{EOM_C_final}), can now be solved using the matrix elements $\eta_{i_{\kappa}}^{j_{\kappa}(\kappa)}(t)$. Thus the EOM for the multispecies TD-RASSCF theory can be solved.
\\
\noindent

\paragraph{General RAS scheme} \hspace{0pt} \label{General RAS scheme}\\

Considering only even excitations provides an efficient and simple way to uncouple the EOM of the multispecies TD-RASSCF theory. Nonetheless, a general RAS scheme, including both even and odd excitations, can be used to construct the total wavefunction. We introduce $N^{(\kappa)}_\text{max}$, the highest number of particles of type $\kappa$ allowed to be excited in the $\mathcal{P}^{(\kappa)}_{2}$ space, such as $N^{(\kappa)}_\text{max}\le N_{\kappa}$. The Fock space spanned by the wavefunction is constructed by considering the successive excitations from $0$ to $N^{(\kappa)}_\text{max}$ particles from the $\mathcal{P}^{(\kappa)}_{1}$-space to the $\mathcal{P}^{(\kappa)}_{2}$-space orbitals. For instance, taking $N^{(\kappa)}_\text{max}=3$, the wavefunction includes all the configurations with $0,1,2$ and $3$ particles of type $\kappa$ in the $\mathcal{P}^{(\kappa)}_{2}$-space orbitals. Generally speaking, the configurational space is decomposed into a direct sum of $N^{(\kappa)}_\text{max}+1$ spaces, 
\begin{equation} 
\mathcal{V}_{\text{RAS}_{\kappa}} = \mathcal{V}^{(\kappa)}_{0}\oplus\mathcal{V}^{(\kappa)}_{1}\oplus \cdots \oplus \mathcal{V}^{(\kappa)}_{N_\text{max}}.
\end{equation}
In this case, the EOM for the amplitudes and the $\mathcal{P}^{(\kappa)}$-space orbitals remain coupled, because $\dot{\rho}_{i'_{\kappa}}^{j''_{\kappa}(\kappa)}(t)\ne0$. Inserting the result obtained for the time-derivative of the amplitudes, Eq. (\ref{EOM_C_final}),  into the expression of time-derivative of the reduced one-body density matrix for the particles of type $\kappa$, Eq. (\ref{TD_rho_K}), we obtain
\begin{multline} 
i\dot{\rho}_{i'_{\kappa}}^{j''_{\kappa}(\kappa)}(t) = \langle\Psi(t)|\left[iD^{(\kappa)}(t)+i\sum_{\gamma\ne\kappa}D^{(\gamma)}(t)-H(t)\right]\Pi(t)\ b_{i'_{\kappa}}^{\dag(\kappa)}b^{(\kappa)}_{j''_{\kappa}}|\Psi(t)\rangle\\
-\langle\Psi(t)|b_{i'_{\kappa}}^{\dag(\kappa)}b^{(\kappa)}_{j''_{\kappa}} \ \Pi(t)\left[iD^{(\kappa)}(t)+i\sum_{\gamma\ne\kappa}D^{(\gamma)}(t)-H(t)\right]|\Psi(t)\rangle,
\end{multline}
with $\Pi(t) = \sum_{\vec{n}_{1},\dots,\vec{n}_{K}\in \cal{V}_\text{RAS}} |\vec{n}_{1},\dots,\vec{n}_{K},t\rangle\langle\vec{n}_{1},\dots,\vec{n}_{K},t|$, the projector onto the $\mathcal{V}_\text{RAS}$ space. Inserting this result in Eq. (\ref{P_space_general_K}) and rearranging the different terms we obtain a new formulation of the $\cal{P}^{(\kappa)}$-space orbital EOM,
\begin{multline} \label{P_space_EOM_proj}
 \langle\Psi(t)|\left[iD^{(\kappa)}(t)+i\sum_{\gamma\ne\kappa}D^{(\gamma)}(t)-H(t)\right]\left[\mathds{1}-\Pi(t)\right]b_{i'_{\kappa}}^{\dag(\kappa)}b^{(\kappa)}_{j''_{\kappa}}|\Psi(t)\rangle\\
 -\langle\Psi(t)|b_{i'_{\kappa}}^{\dag(\kappa)}b^{(\kappa)}_{j''_{\kappa}}\left[\mathds{1}-\Pi(t)\right]\left[iD^{(\kappa)}(t)+i\sum_{\gamma\ne\kappa}D^{(\gamma)}(t)-H(t)\right]|\Psi(t)\rangle=0.
\end{multline}
 For $|\phi^{(\kappa)}_{i'_{\kappa}}\rangle\in \mathcal{P}^{(\kappa)}_{1}$ and $|\phi^{(\kappa)}_{j''_{\kappa}}\rangle\in \mathcal{P}^{(\kappa)}_{2}$, we note that $b_{i'_{\kappa}}^{\dag(\kappa)}b^{(\kappa)}_{j''_{\kappa}}|\Psi(t)\rangle$ belongs to $\mathcal{V}_{\text{RAS}_\kappa}$, with one particle from the $\mathcal{P}^{(\kappa)}_{2}$ orbitals being annihilated and the creation of one particle in the $\mathcal{P}^{(\kappa)}_{1}$ orbitals. It follows that $[\mathds{1}-\Pi(t)]b_{i'_{\kappa}}^{\dag(\kappa)}b^{(\kappa)}_{j''_{\kappa}}|\Psi(t)\rangle= 0$. On the other hand,  $\langle\Psi(t)|b_{i'_{\kappa}}^{\dag(\kappa)}b^{(\kappa)}_{j''_{\kappa}}$, provides configurations with the creation of an additional particle in the $\mathcal{P}^{(\kappa)}_{2}$ orbitals, which can lie in $\mathcal{V}^{(\kappa)}_{N^{(\kappa)}_\text{max}+1}$, not included in $\mathcal{V}_{\text{RAS}_\kappa}$. In this case, $\langle\Psi(t)|b_{i'_{\kappa}}^{\dag(\kappa)}b^{(\kappa)}_{j''_{\kappa}}[\mathds{1}-\Pi(t)]\ne 0$ and  Eq. (\ref{P_space_EOM_proj}) simplifies to,
 \begin{equation}\label{P_space_EOM_proj_simply}
\langle\Psi(t)|b_{i'_{\kappa}}^{\dag(\kappa)}b^{(\kappa)}_{j''_{\kappa}}\left[\mathds{1}-\Pi(t)\right]\left[iD^{(\kappa)}(t)+i\sum_{\gamma\ne\kappa}D^{(\gamma)}(t)-H(t)\right]|\Psi(t)\rangle=0,
 \end{equation}   
 with
  \begin{equation}
Ê\langle\Psi(t)|b_{i'_{\kappa}}^{\dag(\kappa)}b^{(\kappa)}_{j''_{\kappa}}\left[\mathds{1}-\Pi(t)\right]= \sum_{\vec{n}_{\kappa}\in \mathcal{V}^{(\kappa)}_{N^{(\kappa)}_\text{max}}} \sum_{\vec{n}_{1},\dots,\vec{n}_{\kappa-1},\vec{n}_{\kappa+1},\dots,\vec{n}_{K}} C_{\vec{n}_{1},\dots,\vec{n}_{K}}^{*}(t)\langle\vec{n}_{1},\dots,\vec{n}_{K},t|b_{i'_{\kappa}}^{\dag(\kappa)}b^{(\kappa)}_{j''_{\kappa}}.
 \end{equation}   
Equation (\ref{P_space_EOM_proj_simply}) remains coupled to Eq. (\ref{EOM_C_final}) because of the presence of the operators $D^{(\kappa)}(t)$ and $D^{(\gamma)}(t)$ with $\gamma\ne \kappa$. Considering the terms with the operators  $D^{(\gamma)}(t)$ and using its explicit expression we have,
 \begin{multline}
\langle\Psi(t)|b_{i'_{\kappa}}^{\dag(\kappa)}b^{(\kappa)}_{j''_{\kappa}}\left[\mathds{1}-\Pi(t)\right]\left[i\sum_{\gamma\ne\kappa}D^{(\gamma)}(t)\right]|\Psi(t)\rangle= \\
\sum_{\gamma\ne\kappa}\sum_{\vec{n}_{\kappa}\in \mathcal{V}^{(\kappa)}_{N^{(\kappa)}_\text{max}}} \sum_{\vec{n}_{1},\dots,\vec{n}_{\kappa-1},\vec{n}_{\kappa+1},\dots,\vec{n}_{K}} \sum_{p_{\gamma}q_{\gamma}}  iC_{\vec{n}_{1},\dots,\vec{n}_{K}}^{*}(t)\eta_{q_{\gamma}}^{p_{\gamma}(\gamma)}(t)\langle\vec{n}_{1},\dots,\vec{n}_{K},t|b_{i'_{\kappa}}^{\dag(\kappa)}b^{(\kappa)}_{j''_{\kappa}}b_{p_{\gamma}}^{\dag(\gamma)}b^{(\gamma)}_{q_{\gamma}}|\Psi(t)\rangle.
 \end{multline}   
The operators $b_{i'_{\kappa}}^{\dag(\kappa)}b^{(\kappa)}_{j''_{\kappa}}$ acting on the wavefunction $|\Psi(t)\rangle$ annihilate a particle in the orbital $|\phi^{(\kappa)}_{j''_{\kappa}}(t)\rangle$ of $\mathcal{P}^{(\kappa)}_{2}$ and create one in the orbital $|\phi^{(\kappa)}_{i'_\kappa}(t)\rangle$ in $\mathcal{P}^{(\kappa)}_{1}$. Thus the resulting wavefunction does not have any configuration belonging to $\mathcal{V}^{(\kappa)}_{N^{(\kappa)}_\text{max}}$. Moreover, $b_{p_{\gamma}}^{\dag(\gamma)}b^{(\gamma)}_{q_{\gamma}}$ acts only on the particles of type $\gamma \ne\kappa$ and hence does not change configurations of the particles of type $\kappa$. Taking now the inner product with $\langle\vec{n}_{1},\dots,\vec{n}_{K},t|$ leads to zero, because the index $\vec{n}_{\kappa}$ is restricted to $\mathcal{V}^{\kappa}_{N^{(\kappa)}_\text{max}}$. The same results are obtained for the terms $H_{\gamma}(t)$ and $H_{\gamma\mu}(t)$ with $\{\gamma , \mu\}\ne\kappa$ [see Eqs. (\ref{H_intra})-(\ref{H_inter})] as the operators do not act on the particles of type $\kappa$. The $\cal{P}^{(\kappa)}$-space EOM for the particles of type $\kappa$ can be rewritten,
 \begin{equation}
 \langle\Psi|b_{i'_{\kappa}}^{\dag(\kappa)}b^{(\kappa)}_{j''_{\kappa}}\left[\mathds{1}-\Pi(t)\right]\left[iD^{(\kappa)}(t)-H_{\kappa}(t)-\frac{1}{2}\sum_{\gamma\ne\kappa}H_{\kappa\gamma}(t)\right]|\Psi(t)\rangle=0,
 \end{equation}   
or equivalently, using the expression of $H_{\kappa}(t)$ [Eq. (\ref{H_intra})], $H_{\kappa\gamma}(t)$ [Eq. (\ref{H_inter})] and $D^{(\kappa)}(t)$ [Eq. (\ref{D_kappa})],
\begin{multline}\label{final_P_space_general_RAS_K}
\sum_{k''_{\kappa}l'_{\kappa}}\left[i\eta_{l'_{\kappa}}^{k''_{\kappa}(\kappa)}(t)-h_{l'_{\kappa}}^{k''_{\kappa}(\kappa)}(t)\right] \zeta_{k''_{\kappa}i'_{\kappa}}^{l'_{\kappa}j''_{\kappa}(\kappa)}(t) = \frac{1}{2} \sum_{k_{\kappa}l_{\kappa}m_{\kappa}n_{\kappa}}v_{l_{\kappa}n_{\kappa}}^{k_{\kappa}m_{\kappa}(\kappa)}(t)\zeta_{k_{\kappa}m_{\kappa}i'_{\kappa}}^{l_{\kappa}n_{\kappa}j''_{\kappa}(\kappa)}(t)\\
+\frac{1}{2}\sum_{\gamma\ne\kappa}\sum_{k''_{\kappa}k_{\gamma}l'_{\kappa}l_{\gamma}}w_{l'_{\kappa}l_{\gamma}}^{k''_{\kappa}k_{\gamma}(\kappa\gamma)}(t)\zeta_{k''_{\kappa}k_{\gamma}i'_{\kappa}}^{l'_{\kappa}l_{\gamma}j''_{\kappa}(\kappa\gamma)}(t),
\end{multline}
where we introduced the fourth- and six-order tensors,
\begin{eqnarray}
\zeta_{k''_{\kappa}i'_{\kappa}}^{l'_{\kappa}j''_{\kappa}(\kappa)} &=& \langle\Psi(t)|b_{i'_{\kappa}}^{\dag(\kappa)}b^{(\kappa)}_{j''_{\kappa}}\left[\mathds{1}-\Pi(t)\right]b_{k''_{\kappa}}^{\dag(\kappa)}b^{(\kappa)}_{l'_{\kappa}}|\Psi(t)\rangle, \\
\zeta_{k_{\kappa}m_{\kappa}i'_{\kappa}}^{l_{\kappa}n_{\kappa}j''_{\kappa}(\kappa)}&=&\langle\Psi(t)|b_{i'_{\kappa}}^{\dag(\kappa)}b^{(\kappa)}_{j''_{\kappa}}\left[\mathds{1}-\Pi(t)\right]b_{k_{\kappa}}^{\dag(\kappa)}b_{m_{\kappa}}^{\dag(\kappa)}b^{(\kappa)}_{n_{\kappa}}b^{(\kappa)}_{l_{\kappa}}|\Psi(t)\rangle,\\
\zeta_{k''_{\kappa}k_{\gamma}i'_{\kappa}}^{l'_{\kappa}l_{\gamma}j''_{\kappa}(\kappa\gamma)}&=&\langle\Psi(t)|b_{i'_{\kappa}}^{\dag(\kappa)}b^{(\kappa)}_{j''_{\kappa}}\left[\mathds{1}-\Pi(t)\right]b_{k''_{\kappa}}^{\dag(\kappa)}b^{(\kappa)}_{l'_{\kappa}}b_{k_{\gamma}}^{\dag(\gamma)}b^{(\gamma)}_{l_{\gamma}}|\Psi(t)\rangle.
\end{eqnarray}
The EOM of the multispecies TD-RASSCF theory can now be solved using a general RAS scheme. First one solves the $\cal{P}^{(\kappa)}$-space orbital EOM for each type of particles, Eqs. (\ref{final_P_space_general_RAS_K}), to evaluate the upper (or lower) off-diagonal matrix elements of $\eta^{(\kappa)}(t)$. The knowledge of these matrices can then be used to solve the EOM for the amplitudes, Eq. (\ref{EOM_C_final}). The $\cal{Q}^{(\kappa)}$-space orbitals equations, Eqs. (\ref{final_eq_Q_K}), are not coupled to the others and can be evaluate in any order.

It is important to note that the uncoupled EOM have been obtained without any assumption on the value of $N^{(\kappa)}_\text{max}$. Thus, depending of the physical system under investigation it is possible to describe the different types of particles with different accuracy. It can be especially useful for the case of a mixture of fermions and bosons, in which case we can introduce a set of $\mathcal{P}^{(\kappa)}_{0}$ orbitals for the fermions. It is also interesting to note that we derived the EOM for the even excitation RAS scheme in Sec. \ref{Even excitation RAS scheme} and the general RAS scheme in Sec. \ref{General RAS scheme} considering that all kinds of particles are described using either the first or the second RAS scheme. Nonetheless, according to the results obtained, it is also possible to consider only even excitation schemes for some of the particles and general RAS schemes for the remaining types. For this case, we should solve Eq. (\ref{final_P_space_general_RAS_K}) to evaluate the matrix elements of $\eta^{(\kappa)}(t)$ for the particles described by the general RAS scheme and Eq. (\ref{final_P_space_even_K}) for the matrix elements of $\eta^{(\kappa)}(t)$ of particles described by the even excitation RAS scheme. In closing this section, we mention that the EOM for the case of a single type of particles, either bosons or fermions, are contained in the set of EOM presented here as a limiting case.

\section{A critical test case: Asymmetric mixture of bosons}
To validate the theory and to illustrate its properties we choose, as a first application, to consider a challenging ground-state (GS) test case.   We compare the GS energy of a system using different level of accuracy to approximate the wavefunction, and by virtue of the variational principle the more accurate description the lower the energy. For mixtures, we discussed in Sec. \ref{mb_wf} the exponential scaling of the number of configurations $\mathcal{N}_{c}$ with respect to the number of orbitals and particles in FCI approaches. Thus, except for mixtures with an handful of particles of each type, comparison with a full configurational \textit{Ansatz} is not possible and comparison with an exact analytical model is therefore more suitable for validation of the present theory. To this end, we consider the harmonic interaction model (HIM) \cite{Cohen85,Yan03} that was recently extended to mixtures of two types of bosons \cite{Klaiman17_1,Klaiman17_2}. This model constitutes one of the rare cases with an analytical solution for the GS of a many-body system. We restrict our application to a 1-dimensional (1D) system. For cold atoms, a 1D system can be achieved experimentally by a strong transversal confinement of an atomic cloud \cite{Gorlitz01,Moritz03,Laburthe04,Hofferberth07}. In cartesian coordinates and setting $\hbar = 1$,  the Hamiltonian  of the system reads,
\begin{multline}\label{H_appli}
H(x_{1},\dots,x_{N_{1}},y_{1},\dots,{y_{N_{2}}})=\sum_{i=1}^{N_{1}}\left(-\frac{1}{2m_{1}}\frac{\partial^{2}}{\partial x_{i}^{2}}+\frac{1}{2}m_{1}\omega^{2}x_{i}^{2}\right)+\sum_{i=1}^{N_{2}}\left(-\frac{1}{2m_{2}}\frac{\partial^{2}}{\partial y_{i}^{2}}+\frac{1}{2}m_{2}\omega^{2}y_{i}^{2}\right)\\
+\lambda_{1}\sum_{i=1}^{N_{1}}\sum_{j>i}^{N_{1}}\left(x_{i}-x_{j}\right)^{2}+\lambda_{2}\sum_{i=1}^{N_{2}}\sum_{j>i}^{N_{2}}\left(y_{i}-y_{j}\right)^{2}+\lambda_{12}\sum_{i=1}^{N_{1}}\sum_{j=1}^{N_{2}}\left(x_{i}-y_{j}\right)^{2},
\end{multline}       
with $x_{i}$ ($y_{i}$) the coordinates of the particles of type $1$ ($2$). We consider the same trap frequency, $\omega$, for both species, $\lambda_{i}$ ($i=1,2$) is the intra-species interactions strength for the species $i$, $\lambda_{12}$ is the inter-species interaction strength, $N_{i}$ is the number of bosons of the species $i$ of mass $m_{i}$. The intra- and inter-species interactions are attractive for positive values of $\lambda_{i}$ and $\lambda_{12}$ and repulsive otherwise. Using Jacobi coordinates, the Hamiltonian (\ref{H_appli}) is separable in $N=N_{1}+N_{2}$ uncoupled harmonic oscillators with the GS energy given by \cite{Klaiman17_2},
\begin{equation}\label{Ex_GS_En}
E_{ex} = \frac{1}{2}\left[E_{rel}^{1} + E_{rel}^{2} + E_{rel}^{12} + E_{com}\right], 
\end{equation}       	
where,
\begin{equation}
E_{rel}^{1}=(N_{1}-1)\sqrt{\omega^{2}+\frac{2}{m_{1}}\left(\lambda_{1}N_{1}+\lambda_{12}N_{2}\right)},Ê
\end{equation}
\begin{equation}
E_{rel}^{2}=(N_{2}-1)\sqrt{\omega^{2}+\frac{2}{m_{2}}\left(\lambda_{2}N_{2}+\lambda_{12}N_{1}\right)},Ê
\end{equation}
\begin{equation}
E_{rel}^{12}=\sqrt{\omega^{2}+2\lambda_{12}\left(\frac{N_{1}}{m_{2}}+\frac{N_{2}}{m_{1}}\right)}, 
\end{equation}
\begin{equation}
E_{com} = \omega. 
\end{equation}       
Here $E_{rel}^{i}$, with $i=1,2$, is the energy of the intra-species relative motion with respect to the center-of-mass (\textit{c.o.m.}), $E_{rel}^{12}$ is the energy of the inter-species relative coordinate, i.e., between the \textit{c.o.m.} of species 1 and the \textit{c.o.m.} of species 2, and $E_{com}$ is the energy in the \textit{c.o.m.} coordinate. 

In the following numerical simulations, we consider an asymmetric number of particles of the two species and coupling strengths chosen to mimic the interaction of a few impurities with an ideal non-interacting Bose gas. Specifically, we consider $N_{1}=100$ bosons of type 1 and $N_{2}=4$ bosons of type 2 with identical masses, $m_{1}=m_{2}=1$, set to unity. We consider a harmonic trapping potential with a frequency $\omega=1$. The bosons of type 1 are non-interacting, i.e., $\lambda_1=0$, while the bosons of type 2 experience an interaction of strength $\lambda_{2}=0.5$. The two types of bosons interact with a strength $\lambda_{12}=0.1$. The analytically exact energy of the system is $E_{ex}=76.7457424377$, determined from Eq. (\ref{Ex_GS_En}). To obtain the GS energy using the multispecies TD-RASSCF theory, we propagate the EOM in imaginary time, i.e., $t\rightarrow -i\tau$ with $\tau \in \mathds{R}$, such that starting from an initial guess the wavefunction converges to the GS of the Hamiltonian \cite{Kosloff86}. The time-dependent orbitals are expanded over a spatial range from $[ - 5; 5]$ in a time-independent primitive basis consisting of 101 quadrature points from a sine-DVR \cite{Light85,Beck00}. The convergence threshold is set such that the energy difference between two successive 
time-steps
should be below $<10^{-13}$. The numerical integration of the EOM is performed using the Adams-Bashforth-Moulton (ABM) predictor-corrector integrator \cite{Beck00,MCTDH}.

\begin{figure}
  \centering
    \includegraphics[scale=0.3]{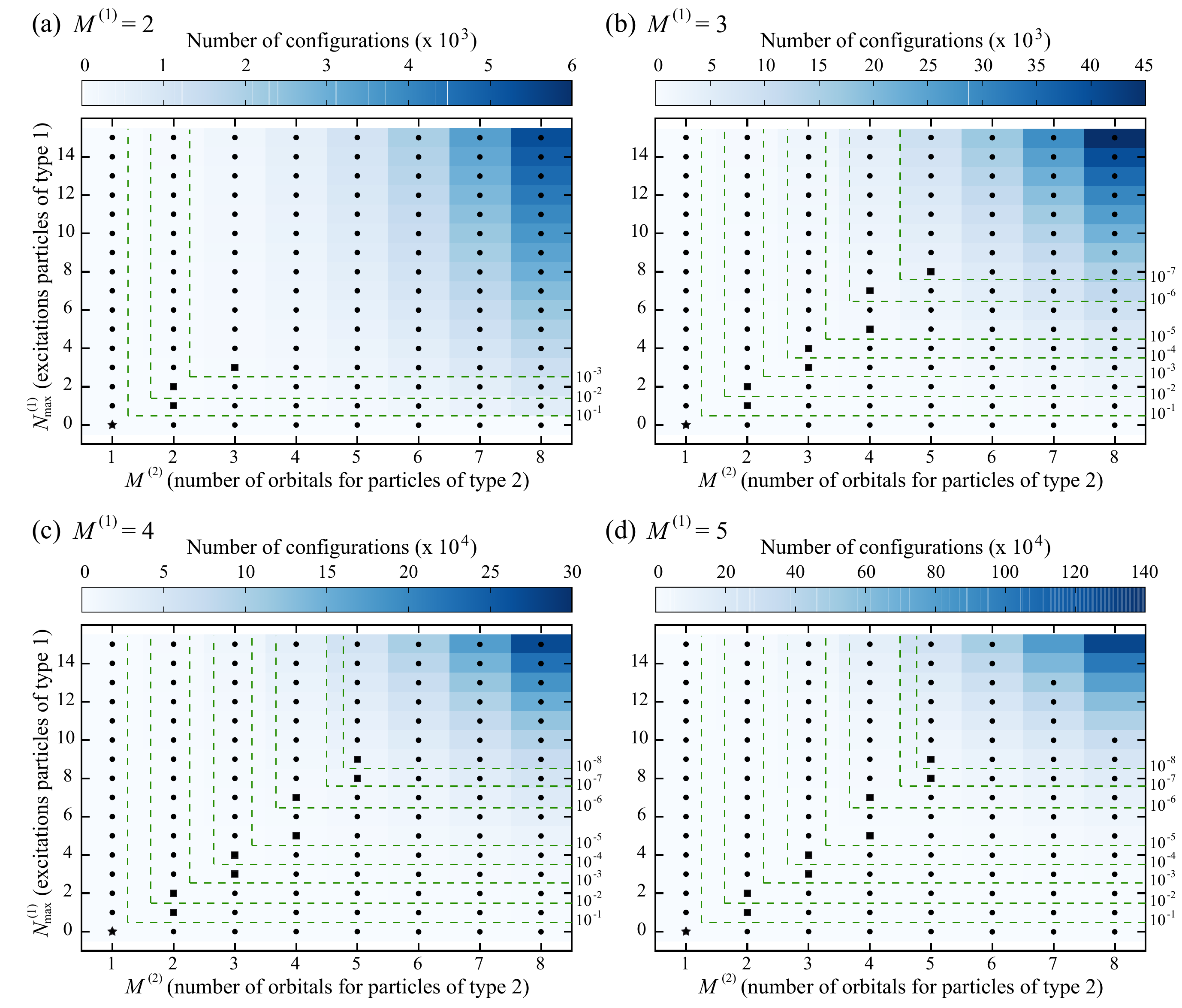}
\caption{ 
Error of the GS energy, $\Delta E$, obtained with the TD-RASSCF-B method using different RAS schemes in comparison to the analytical energy, see Eq. (\ref{DeltaE}). 
The results were obtained from a relaxation with the Hamiltonian defined in Eq. (\ref{H_appli}) with two types of bosons, $N_{1}=100$, $N_{2}=4$, $\lambda_{1}=0$, $\lambda_{2}=0.5$ and $\lambda_{12}=0.1$. (a) Results obtained using $M^{(1)}=2$ orbitals for particles of type 1 with $M_{1}^{(1)}=1$ and $M_{2}^{(1)}=1$, (b) using $M^{(1)}=3$ orbitals for particles of type 1 with $M^{(1)}_{1}=1$ and $M^{(1)}_{2}=2$, (c)  using $M^{(1)}=4$ orbitals for particles of type 1 with $M^{(1)}_{1}=1$ and $M^{(1)}_{2}=3$ and (d) using $M^{(1)}=5$ orbitals for particles of type 1 with $M^{(1)}_{1}=1$ and $M^{(1)}_{2}=4$. The 4 particles of type 2 are described using all configurations, i.e., by a FCI MCTDHB \textit{Ansatz} and the horizontal axis indicates the number of orbitals used, while the vertical axis indicates the number of excitations,  $N_\text{max}^{(1)}$, allowed from the $\mathcal{P}^{(1)}_{1}$ to the $\mathcal{P}^{(1)}_{2}$ orbitals for the 100 particles of type 1. Note that for $M^{(2)}=1$ and no excitation for the particles of type 1, i.e., $N_\text{max}^{(1)} =0$ the result is equivalent to the one obtain by two coupled GP equations. The (black) dots indicate the performed simulations, the dashed (green) lines indicate isocontours for the energy difference between the exact energy ($E_{ex}=76.7457424377$) and the energies obtained from the simulations and the color map gives the number of the configurations. 
The stars in the lower left corners denote the GP results. The squares denote the results obtained for a certain accuracy with the smallest number of configurations. 
}   \label{Fig_3} 
 \end{figure}

	 The multispecies TD-RASSCF theory offers a large flexibility to approximate the total wavefunction of the system using different levels of accuracy for the different types of particles. Specifically, for the system at hand, we consider all configurations, i.e., equivalent to the MCTDHB \textit{Ansatz}, for the 4 bosons of type 2. Thus, for this type of particles we remain with only one parameter to control the accuracy: The total number of orbitals $M^{(2)}$. Note that using this FCI wavefunction for the particles of type 2 is equivalent to a choice of a RAS scheme with only $\mathcal{P}^{(2)}_{1}$-space orbitals, i.e., $M^{(2)}=M^{(2)}_{1}$ and $M^{(2)}_{0}=M^{(2)}_{2}=0$, see Fig. \ref{Fig_2}. This choice of describing the particles of type 2 by the FCI expansion is motivated by the fact that the particles interact with each other with a relatively large interaction strength. In the results for the GS energy presented in Fig. \ref{Fig_3} we consider $1\le M^{(2)}\le 8$, with $M^{(2)}=1$ being equivalent to the mean-field GP \textit{Ansatz} for the particles of type 2. Concerning the 100 bosons of type 1, the $\mathcal{P}^{(1)}$-space orbitals $M^{(1)}$ are shared between the $\mathcal{P}^{(1)}_{1}$- and the $\mathcal{P}^{(1)}_{2}$-space with $M^{(1)}_{1}=1$ orbital and $M^{(1)}_{2}=M^{(1)}-M^{(1)}_{1}$ orbitals, respectively. The results of Fig. \ref{Fig_3} were obtained with $M^{(1)}=2$ in panel (a), $M^{(1)}=3$ in panel (b), $M^{(1)}=4$ in panel (c) and $M^{(1)}=5$ in panel (d). The last choice of  parameter to fully define the wavefunction is the excitation scheme from the $\mathcal{P}^{(1)}_{1}$- to the $\mathcal{P}^{(1)}_{2}$-space for the particles of type 1. We use the general RAS scheme (Sec. \ref{General RAS scheme}), i.e., fixing the highest excitation allowed, $N_\text{max}^{(1)}$, all successive excitations from $0$ to $N_\text{max}^{(1)}$ are included in the wavefunction. In Fig. \ref{Fig_3}, the results were obtained for $0\le N_\text{max}^{(1)}\le15$, with $N_\text{max}^{(1)}=0$ being equivalent to the mean-field GP \textit{Ansatz} for the particles of type 1. Note that this specification is not possible in the (ML-)MCTDHB framework, where all excitations in the orbital space are included.  
	  	 
	  To compare the GS energy obtained with the different \textit{Ans\"atze} of the total wavefunction, we report in Fig. \ref{Fig_3} the accuracy (dashed (green) lines), 
\begin{equation}\label{DeltaE}
\Delta E = E_{approx}-E_{ex},
\end{equation}
 with $E_{approx}$ the energy resulting from the evaluation of the EOM for the multispecies TD-RASSCF theory for the various \textit{Ans\"atze} and the exact energy $E_{ex}$ from Eq. (\ref{Ex_GS_En}). The GP result for both types of particles is obtained for $M^{(2)}=1$ and $N_\text{max}^{(1)}=0$ (left corner at the bottom of each panel), and is the least accurate results with $E_{GP}=76.8799982961$ and $\Delta E =0.134$. From Fig. \ref{Fig_3} (a) we see that for $M^{(1)}=2$ we obtain $\Delta E>10^{-4}$, irrespectively of the number of $M^{(2)}$ orbitals for the particles of type 2 and irrespectively of the value of $N_\text{max}^{(1)}$ the highest excitations allowed for the particles of type 1. The convergence is obtained for $M^{(2)}\ge 3$ and $N_\text{max}\ge 3$ with 60 configurations, in the sense that no significative improvement of the energy is obtained when $M^{(2)}$ or $N_\text{max}^{(1)}$ increase further. We confirm this finding by increasing $N_\text{max}^{(1)}$ to $25$ and we found that  $\Delta E$ remains above $10^{-4}$. Thus, using only $M^{(1)}=2$ orbitals is not sufficient to converge to the exact result but provides a significative improvement of the accuracy in comparison to the mean-field GP result. It should be stressed that the RAS scheme with only 60 configurations provides similar accuracy as a FCI MCTDHB \textit{Ansatz} including at least 1515 configurations for 2 orbitals for particles of type 1 and 3 orbitals for particles of type 2, see Table I. 

	The results of Fig. \ref{Fig_3}(a) show that using $M^{(1)}=2$ for the particles of type 1 is not sufficient to an accurate description of the system. It is necessary to include $M^{(1)}=3$ orbitals, Fig. \ref{Fig_3}(b), for the particles of type 1  to obtain accurate results for the GS energy. In Fig. \ref{Fig_3}(b) we see that $10^{-8}<\Delta E<10^{-7}$ is obtained for $M^{(2)}\ge5$ and $N^{(1)}_\text{max}\ge8$ and increasing $N^{(1)}_\text{max}$ to 25 does not change the accuracy of the results. The best accuracy, i.e., $\Delta E<10^{-7}$, can be obtained at a minimal cost using $M^{(2)}=5$ and $N^{(1)}_\text{max}=8$, leading to $3 150$ configurations. In comparison,  $360 570$ configurations are obtained without a restriction on the configurational space of particles of type 1. The multispecies TD-RASSCF theory proves that only few excitations, $N^{(1)}_\text{max}\ge8$, are sufficient to describe the correlation between the two types of particles, while the wavefunction includes $3 150$ configurations. 
	
	Increasing the number of $M^{(1)}$ orbitals to $4$, Fig. \ref{Fig_3}(c), the accuracy becomes better with $10^{-9}<\Delta E<10^{-8}$ and the same accuracy is obtained when the number of orbitals is increased to $M^{(1)}=5$ [Fig. \ref{Fig_3}(d)]. Accuracy below $10^{-9}$ is not obtained when the number of quadrature points is increased to $201$ or by using Runge-Kutta to integrate the EOM, thus the GS energy has converged with respect to the parameters of the wavefunction. Convergence is obtained at a minimal cost for the parameters $N^{(1)}_\text{max}=9$, $M^{(1)}=4$ and $M^{(2)}=5$ and $15400$ configurations, and this number could be reduced by considering a RAS scheme for the particles of type 2. Without restriction on the configurational space of the particles of type 1, the wavefunction includes $12379570$ configurations, i.e., $800$ times more than needed to converge to the numerically exact result. In this example we know an analytically exact solution for the GS energy, but usually this energy is not known. Nonetheless, the application illustrates the strength of the multispecies TD-RASSCF method since this method by its own can ensure the convergence of the calculations by varying the different parameters without an exponential increase of the number of configurations. This latter possibility is usually not present when the FCI space is used. If the two types of particles are described with an MCTDHB \textit{Ansatz}, to ensure that the calculation converged with $M^{(1)}=4$ and $M^{(2)}=5$ orbitals, a calculation with $M^{(1)}=5$ and $M^{(2)}=6$ orbitals should be perform, but the wavefunction includes $579363876$ configurations, beyond what can be presently handled computationally.    
 \begin{table}[h!]
 \fontsize{8}{10}\selectfont
\caption{Ground-state energy of $100$ bosons of type 1 and $4$ bosons of type 2 trapped in a 1D harmonic potential with harmonic inter-particle interaction strengths $\lambda_{1}=0$ and $\lambda_{2}=0.5$ and the harmonic intra-interaction strength $\lambda_{12}=0.1$ [see Eq. (\ref{H_appli})]. The multispecies TD-RASSCF calculations were performed using the general RAS scheme, see Sec. \ref{General RAS scheme}. A single ${\cal P}^{(1)}_{1}$ orbital, $M^{(1)}_{1}=1$, and $M^{(1)}_{2}=M^{(1)}-1$ ${\cal P}^{(1)}_{2}$ orbitals are used for the particles of type 1. The 4 particles of type 2 are described using all configurations, i.e., by an MCTDHB \textit{Ansatz} with different numbers of orbitals $M^{(2)}$ from $1$ to $5$. We report, for a certain accuracy, the results obtained with the smallest number of configurations in the wavefunction expansion, see Eq. (\ref{RAS_wf}), extracted from Fig. \ref{Fig_3} and represented by full squares or stars. $N_\text{max}^{(1)}$ denotes the maximum excitation number for the 100 particles of type 1. $\mathcal{N}_{c}$ denotes the number of configurations for the different types of calculations considered.  The entries in the FCI results of the lower right corner are empty because of the intractable size of the configurational space.}
  \begin{tabular}{p{3cm} p{0.1cm} p{3cm} p{3cm} p{3cm} p{3cm}  p{0.1cm}} 
  \hline \hline
  \multicolumn{1}{c}{Accuracy } & \multicolumn{5}{c}{ $M^{(1)}$ orbitals } &    \\
       \cline{2-7}
    \centering  & &\centering $2$ &\centering $3$  &\centering $4$ &\centering $5$  &\\
    \centering  & &\centering  $N_{\text{max}}^{(1)}\text{ }|\text{ }M^{(2)}\text{ }|\text{ }\mathcal{N}_{c}$ &\centering $N_{\text{max}}^{(1)}\text{ }|\text{ }M^{(2)}\text{ }|\text{ }\mathcal{N}_{c}$  &\centering $N_{\text{max}}^{(1)}\text{ }|\text{ }M^{(2)}\text{ }|\text{ }\mathcal{N}_{c}$ &\centering $N_{\text{max}}^{(1)}\text{ }|\text{ }M^{(2)}\text{ }|\text{ }\mathcal{N}_{c}$  &\\
    
    \hline 
    \centering $<1$ && \centering $76.8799982961$ &  \centering $76.8799982961$ &  \centering $76.8799982961$  &  \centering  $76.8799982961$   & \\
         && \centering $0\ |\ 1\ |\ 1$ & \centering $0\ |\ 1\ |\ 1$ & \centering  $0\ |\ 1\ |\ 1$  &  \centering  $0\ |\ 1\ |\ 1$   & \\
    \centering $<10^{-1}$ && \centering $76.7630671046$ & \centering $76.7630389012$ & \centering   $76.7630388981$  &  \centering  76.7630388981   & \\
         && \centering $1\ |\ 2\ |\ 10$ & \centering $1\ |\ 2\ |\ 15$ & \centering  $1\ |\ 2\ |\ 20$  &  \centering  $1\ |\ 2\ |\ 25$   & \\ 
    \centering $<10^{-2}$ && \centering $76.7503076208$ & \centering $76.7502536661$ & \centering   $76.7502536571$  &  \centering  76.7502536571  & \\
         && \centering $2\ |\ 2\ |\ 15$ & \centering $2\ |\ 2\ |\ 30$ & \centering  $2\ |\ 2\ |\ 50$  &  \centering  $2\ |\ 2\ |\ 75$   & \\ 
    \centering $<10^{-3}$ && \centering $76.7462653876$ & \centering $76.7461619352$ & \centering   $76.7461618891$  &  \centering  76.7461618891   & \\
         && \centering $3\ |\ 3\ |\ 60$ & \centering $3\ |\ 3\ |\ 150$ & \centering $3\ |\ 3\ |\ 300$   &  \centering $3\ |\ 3\ |\ 525$  & \\ 
    \centering $<10^{-4}$ && \centering - & \centering  $76.7458263013$ & \centering   $76.7458262476$  &  \centering  76.7458262476  & \\
         && \centering  & \centering $4\ |\ 3\ |\ 225$ & \centering  $4\ |\ 3\ |\ 525$ &  \centering  $4\ |\ 3\ |\ 1050$  & \\      
    \centering $<10^{-5}$ && \centering - & \centering  $76.7457506111$ & \centering   $76.7457505481$  &  \centering  76.7457505481  & \\
         && \centering  & \centering $5\ |\ 4\ |\ 735$ & \centering $5\ |\ 4\ |\ 1960$ &  \centering  $5\ |\ 4\ |\ 4410$   & \\ 
    \centering $<10^{-6}$ && \centering - & \centering  $76.7457429381$ & \centering   $76.7457428741$  &  \centering  76.7457428740  & \\
         && \centering  & \centering $7\ |\ 4\ |\ 1260$ & \centering $7\ |\ 4\ |\ 4200$ &  \centering  $7\ |\ 4\ |\ 11550$   & \\ 
    \centering $<10^{-7}$ && \centering - & \centering  $76.7457425228$ & \centering   $76.7457424584$  &  \centering  76.7457424584   & \\
         && \centering  & \centering $8\ |\ 5\ |\ 3150$ & \centering $8\ |\ 5\ |\ 11550$ &  \centering  $8\ |\ 5\ |\ 34650$   & \\ 
    \centering $<10^{-8}$ && \centering - & \centering - & \centering   $76.7457424430$  &  \centering  76.7457424430  & \\
         && \centering  & \centering  & \centering $9\ |\ 5\ |\ 15400$ &  \centering  $9\ |\ 5\ |\ 50050$ & \\                
    \hline
 \centering FCI && \centering 76.7458841897 & \centering  $76.7457425051$ & \centering   $-$  &  \centering  -   & \\
     && \centering  $\text{all }Ê|\ Ê3\ |\ 1515$ & \centering  $\text{all }Ê|\ Ê5\ |\ 360570$ & \centering $\text{all } |\ 5\ |\ 12379570$ &  \centering  $\text{all }  |\ 5\ |\ 321868820$ & \\   
    \hline \hline
 \end{tabular}
 \end{table}
	
	The multispecies TD-RASSCF method provides a simple way to analyze the role of the inter- and intra-particle correlation, thanks to the clear hierarchy of the RAS scheme.
The numerically exact energy of the GS is obtained for $M^{(1)}=4$, with $M^{(1)}_{1}=1$ and $M^{(1)}_{2}=3$, $N^{(1)}_{\text{max}} = 9$ and $M^{(2)}=5$. The wavefunction includes configurations with at least $91$ bosons occupying the same orbital, thus the particles of type 1 are mainly condensed. Moreover, for $N^{(1)}_{\text{max}} = 9$, we see from Fig. \ref{Fig_3} that increasing the number of orbitals $M^{(1)}$ from $1$ to $4$ substantially decreases the energy of the GS, thus the few particles out of the condensate favorably occupy different orbitals rather than a collective occupation of the orbitals, which explain the small number of excitations required to converge to the exact result. The comparison of the results between the different RAS wavefunctions show that while most particles remain in the motional ground-state orbital, it is important to account for the relatively fast motion of a few particles by including several orbitals, while the highly excited configurations play virtually no role. This result is reminiscent to the one obtained for one type of particles \cite{Leveque17} for interacting bosons, while here the bosons only interact with the second type of particles. The results of Fig. \ref{Fig_3}, show that the mean-field GP description of mixtures cannot be expected to be accurate ($\Delta E > 10^{-1}$).
Interestingly, the results also show that combining a mean-field GP and an MCTDHB \textit{Ansatz} does not improve the results in comparison to the mean-field description (lower stars and dots in Fig. \ref{Fig_3}, $N_\text{max}^{(1)}=0$). The contour lines of Fig. \ref{Fig_3}(b) show that a GP \textit{Ansatz} for particles of type 1 combined with an MCTDHB \textit{Ansatz} for the particles of type 2 only provide poor accuracy with $\Delta E >10^{-1}$, and this even if the particles of type 1 do not interact with each other. The multispecies TD-RASSCF theory proves that only few excitations, $N^{(1)}_\text{max}\ge 9$, are sufficient to describe the correlation between the two types of particles, while the wavefunction includes $15400$ configurations. The number of configurations is much lower than the $12379570$ configurations obtained with an MCTDHB \textit{Ansatz} with $4$ orbitals for the particles of type 1 and $5$ orbitals for the particles of type 2.  
	
\section{Conclusion and Outlook}

In this work, we discussed the multispecies TD-RASSCF theory, an \textit{ab initio} method based on the time-dependent variational principle, using a restricted-active-space and time-dependent orbitals. The equations of motion for the wavefunction \textit{Ansatz} were derived for an arbitrary number of particle types and an arbitrary number of bosonic and fermonic species. Two specific RAS schemes were described to solve the coupled equations of motion  for a given species (i) a scheme with only even excitations and (ii) a scheme with all successive excitations up to a certain cut-off. For both schemes the maximum excitation level can be fixed arbitrarily. The choice of one of these schemes is specific for each species and different species can be treated by different schemes and different maximum excitation levels. Thus the wavefunction \textit{Ansatz}  of the multispecies TD-RASSCF theory is versatile and can be adapted to the physical system at hand. Specifically, this theory bridges the mean-field approximation and the full-configuration interaction \textit{Ansatz} for mixtures of fermions and/or bosons and can, by adjusting the level of approximation, be used to capture the main part of the correlation between particles of each species and between the species. Moreover, as discussed in connection with the derivation of the equations of motion, the single-component case is a limiting case of the multispecies TD-RASSCF theory. 
	
	The combination of time-dependent orbitals and the RAS tackles efficiently and accurately the exponential scaling of the number of configurations.  The theory is variational, thus, for a given number of orbitals, increasing the number of allowed excitations always provides a more accurate description of the time-dependent wavefunction. In this way the accuracy of the simulations can be checked for convergence, which is an important feature for \textit{ab initio} theory. Moreover, the control of the number of configurations obtained through specification of the RAS provides the possibility to include a large number of orbitals in comparison to other available time-dependent wavefunction based methods \cite{Alon08,Cao13,Kronke13}. The multispecies TD-RASSCF theory therefore provides a useful tool to investigate static and dynamic properties of mixtures of particles. 
The explicit time-dependence of the theory, and the RAS scheme with its freedom to explore the important parts of configurational space means that the method in the future can target nonequilibrium dynamics. In this work, as a first application, we investigated the ground-state energy of a mixture of two types of bosons within the harmonic interaction model, for which the exact ground-state energy is known analytically, and which could be used in a  critical test of the theory. We focused on the case of an ideal Bose gas interacting with few relatively stronger interacting impurities. This numerical example provided a convenient way to benchmark the theory, and to illustrate its accuracy. By using different RAS schemes, the role of the correlation in Bose-Bose mixtures can be comprehensively addressed. We showed that the number of configurations has only a small effect on the correlation energy while the number of orbitals and the excitation level plays a major role. The convergence could not be checked without the RAS, i.e., the configurational space in the MCTDHB for mixtures was too large to allow a calculation. This illustrated the strength of the method and the physical result we found means that even for a small depletion of BECs, it is favorable for the particles out of the condensed orbital to occupy higher-energy orbitals separately rather than collectively. The example showed that  the mean-field description of the ideal Bose gas breaks down because of the interaction with the impurity atoms, and this irrespectively of the number of orbitals used to describe these impurities. The small depletion mediated by the impurities has a large impact on the ground state energy and both species must be described beyond the mean-field \textit{Ansatz} to sensitively reduce the error. 	In view of the properties of the multispecies TD-RASSCF theory and its performance for the present ground-state studies, we expect that the theory will be well-suited for the consideration of Bose polaron formation and dynamics \cite{Shchadilova16,2017arXiv171103478G}, also for strong interaction  beyond  the Fr\"ohlich regime/model \cite{Mahan00}; a regime which can be accessed experimentally through Feshbach resonances \cite{Hu16,Jorgensen16}.
	
	For future developments of the theory, it is interesting to note that the multispecies TD-RASSCF theory and the multilayer MCTDH theory for indistinguishable particles \cite{Cao13,Kronke13} tackle the problem of the large number of configurations in different ways. In the first case, the configurations are selected at the level of a single species with the choice of a species-specific RAS scheme. The selected configurations are then used to construct the total wavefunction. Thus, the correlation is approximated at the single-species level and subsequently gives an approximation of the correlation between the different species. In the second case, the entire single-species configuration spaces are used to build an approximation of the correlation between the different species and at the single species-level no approximation is made and consequently a large number of configurations are used. Thus, formulating a theory using a multilayer expansion to approximate the inter-species correlation and a RAS at the single-species level may open a way to investigate dynamics of larger systems with time-dependent wavefunction-based methods.	
	
\section*{Acknowledgments}
We thank Kristian Knakkergaard Nielsen for useful discussions. We acknowledge support from the Villum-Kann Rasmussen (VKR) center of excellence QUSCOPE - Quantum Scale Optical Processes. The numerical results were obtained at the Centre for Scientific Computing, Aarhus.

\bibliography{biblio}

\end{document}